\documentclass[useAMS,usenatbib]{mn2e}
\pdfoutput=1
\usepackage{graphicx}%
\usepackage[fleqn]{amsmath} 
\usepackage{bm}
\usepackage{natbib} 
\usepackage{url}%
\usepackage{hyperref}
\usepackage{gensymb}%
\usepackage{color} 
\usepackage{tabularx} 
\usepackage{xr} 

\definecolor{refcol}{rgb}{0.3,0.,0.4}

\hypersetup{
  colorlinks=true,
  citecolor=refcol,
  filecolor=refcol,
  linkcolor=refcol,
  urlcolor=refcol,
  pdfborder=0 0 0.2
}

\title[Tests of a $C_n^2$ Profiler for the AOF]{Validation Through Simulations of a $C_n^2$ Profiler for the ESO/VLT Adaptive Optics Facility}
\author[A. Garcia-Rissmann et al]{A. Garcia-Rissmann$^{1}\thanks{E-mail:
aureag2014@gmail.com}$, A. Guesalaga$^{2}$, J. Kolb$^{1}$, 
M. Le Louarn$^{1}$, 
P.-Y. Madec$^{1}$, \newauthor B. Neichel$^{3}$ \\
$^{1}$European Southern Observatory, Karl-Schwarzschild-Str. 2, 85748, Garching bei M\"unchen, Germany\\
$^{2}$Pontif\'{i}cia Universidad Cat\'olica de Chile, 4860 Vicu\~na Mackenna, Casilla 7820436, Santiago, Chile\\
$^{3}$Aix Marseille Universit\'e, CNRS, LAM (Laboratoire d'Astrophysique de Marseille) UMR 7326, 13388, Marseille, France}

\begin{document}

\date{Accepted .... Received ...; in original form ...}

\pagerange{\pageref{firstpage}--\pageref{lastpage}} \pubyear{2002}

\maketitle

\label{firstpage}

\begin{abstract}
The Adaptive Optics Facility (AOF) project envisages transforming one of the VLT units into an adaptive telescope and providing its ESO (European Southern Observatory) second generation instruments with turbulence corrected wavefronts. For MUSE and HAWK-I this correction will be achieved through the GALACSI and GRAAL AO modules working in conjunction with a 1170 actuators Deformable Secondary Mirror (DSM) and the new Laser Guide Star Facility (4LGSF).  Multiple wavefront sensors will enable GLAO and LTAO capabilities, whose performance can greatly benefit from a knowledge about the stratification of the turbulence in the atmosphere. This work, totally based on end-to-end simulations, describes the validation tests conducted on a $C_n^2$ profiler adapted for the AOF specifications. Because an absolute profile calibration is strongly dependent on a reliable knowledge of turbulence parameters $r_0$ and $L_0$, the tests presented here refer only to normalized output profiles. Uncertainties in the input parameters inherent to the code are tested as well as the profiler response to different turbulence distributions. It adopts a correction for the unseen turbulence, critical for the GRAAL mode, and highlights the effects of masking out parts of the corrected wavefront on the results. Simulations of data with typical turbulence profiles from Paranal were input to the profiler, showing that it is possible to identify reliably the input features for all the AOF modes. 

\end{abstract}


\section{Introduction}

A reliable knowledge of the atmospheric turbulence distribution allows for the setting of constraints on the design of new adaptive optics (AO) systems, for optimizing their performance and for making a prior assessment of their correction limitations. Such statements are true for AO modes such as laser tomography or multiobject adaptive optics (LTAO and MOAO, respectively) which require an optimization of the wavefront reconstruction along one or more determined lines of sight. In single-,  multiconjugated and ground layer adaptive optics (SCAO, MCAO and GLAO, respectively), the knowledge of the turbulence distribution enables one to determine the level of correction expected for the system, its anisoplanatic behaviour and the residual error that will impact as a degradation of the science point-spread function (PSF) in the telescope focal plane.

Many techniques exist for acquiring information on the turbulence, such as balloon borne experiments to measure temperature structure coefficients, SCIDAR, SLODAR, MASS-DIMM, etc \citep{2014SPIELombardi}. Monitoring programmes have been carried out in Paranal for several years to increase the database that will enable to set a programme to optimize operations -- see e.g. 
\cite{Masciadri:2013aa} -- and to provide parameter constraints for the next generation of VLT and even E-ELT instrumentation \citep{ao4elt3_18787,Sarazin:2013aa}. 

The Adaptive Optics Facility (AOF) is an AO-oriented upgrade to be commissioned in one of the telescope units of the Paranal Observatory in 2016. It counts on the largest Deformable Secondary Mirror (DSM) ever built (1170 voice-coil actuators distributed behind a thin shell of 1120 mm of diameter) and on four Sodium laser side-launch telescopes which are part of the new Laser Guide Star Facility. The AOF will operate in 3 distinct modes (SCAO, GLAO \& LTAO) in accordance to the instruments attached to the 2 Nasmyth ports (GRAAL+HAWK-I, GALACSI+MUSE) and to the Cassegrain port (ERIS) of the telescope. Each of the four Shack-Hartmann wavefront sensors counts on a 40$\times$40 squared subaperture (SA) array, out of which 1240 SAs are considered valid and 1152 are fully illuminated (still neglecting the presence of spiders). The high-order loop is expected to run at 1 kHz. 

The AOF project undergoes in 2014-2015 a phase of laboratory tests with the aid of the dedicated ASSIST bench \citep{Arsenault:2013aa, Arsenault:2014aa}. The main idea is to test and to optimize the system before its delivery for the commissioning in Paranal. For this purpose, calibration and control tools have been developed in order to assist these tests with respect to e.g. DM/WFS mis-registration and turbulence diagnostics \citep{2012SPIE.8447E..5UK,2013aoel.confE...9G, AGR2014}, and are envisaged to be implemented as routines in the real-time computer (RTC) platform SPARTA \citep{2012SPIE.8447E..2QS}. They are planned for use in either real-time or in post-processing modes, with the aim to improve the system performance by e.g. uploading optimized control matrices during operations \citep{2012SPIE.8447E..5UK}. 

Through end-to-end system simulations, this work focused on testing the reliability of a turbulence profiler in several aspects, which includes mimicking the observational conditions in Paranal. 
Section \ref{sec:cn2profiler} describes the profiler code, how it operates and discusses a parameterisation used for the calibration of the profile. In section \ref{sec:tests} we address some sensitivity tests for the assessment of the profiler performance, and section \ref{sec:allmodes} describes its application to a more realistic scenario for all 3 modes. It is worth to point out that parameter dependencies such those on WFS flux and chosen centroid algorithms, although briefly studied as well \citep{AGR2014}, are not addressed in this paper due to them being particularly inherent to the AO system performance, not to the profiler itself; this topic awaits a more dedicated study in the context of the overall quality of the AOF correction. Section \ref{sec:discussion} summarizes and draws conclusions of the main results obtained in this work.

\section[]{THE $C_n^2$ PROFILER}
\label{sec:cn2profiler}
Multi-WFS systems such as GRAAL and GALACSI were designed to probe the atmospheric turbulence in 4 different directions. The common volume of turbulence encompassed by pairs of these sensing directions makes the AOF -- besides its main role in the AO correction -- a suitable tool to trace the vertical turbulence distribution. The 40$\times$40 SA$^2$ array of each wavefront sensor, covering a pupil of 8 m across, can provide an unprecedented combined altitude resolution for this kind of application.
In Open Loop (OL) regime, the 9920 measured slopes can be straightforwardly used to feed a turbulence profiler algorithm. In Closed Loop (CL) regime, Pseudo-Open Loop (POL) slopes for the profiler input can be obtained by multiplying the DM commands ($\bm{c}$) by the interaction matrix ($\mathsf{\mathbf{M}}$) and adding this product to the residual slopes ($\bm{s}_\mathrm{res}$):
\begin{equation}
\label{eq:POL}
\bm{s}_{\mathrm{POL},k} = \bm{s}_{\mathrm{res},k} +  \mathsf{\mathbf{M}} . \bm{c}_{k-\Delta k}
\end{equation}
where the subscript $k$ refers to the frame number and $\Delta k$ to the loop delay, i.e. the delay of the system between the application of the DM command and the availability of the new data from the WFS. This delay lies between 2 and 3 frames for the AOF GLAO and LTAO modes. The interaction matrix $\mathsf{\mathbf{M}}$ for these modes simply replicates the SCAO interaction matrix according to the number of WFSs. The following steps -- valid for both OL and POL slopes -- consist in a subtraction of the temporal averaged slopes and a frame-by-frame removal of the tip-tilt (TT) component from individual WFSs.

Once the OL or POL slopes are correctly computed we can make use of some techniques to reconstruct the $C_n^2$ profile.
These are post-processing tools that have been developed and adapted from the conventional monitoring ones for the reconstruction of the turbulence distribution, based on AO RTC data. For instance,  
 \cite{2008ApOpt..47.1880W} took advantage of the Palomar multiple guide star unit, cross-correlating its WFS data to obtain the distribution of the $C_n^2$ profile as well as wind information. Later, \cite{2012MNRAS.427.2089C} proposed two methods to obtain turbulent profile distributions for the Gemini MCAO system GeMS \citep{Rigaut2014,Neichel2014}. The two algorithms are called `modified-SLODAR' and `wind-profiler', the latter based on Wang et al's work but adding the generalisation to laser beams. These methods have been re-baptized recently as `matrix inversion' (MI) and `Fourier deconvolution' (FD) algorithms, respectively \citep{Valenzuela2014}.

The aforementioned GeMS codes have been kindly provided  to ESO in the form of {\sc Matlab} scripts; they have been adapted and modified for the AOF case to account for its geometry and to provide an alternative method for the unseen turbulence correction. After a phase of understanding the inner workings of the algorithms, only the wind-profiler one was kept, based on the robustness of the results obtained at the higher and the lower ends of the altitude range.  This work is focused on testing this specific algorithm, and we restrict ourselves only to zero-delayed measurements (i.e. no wind speed estimation is attempted). For a recent comparison of both methods applied to  the AOF and a different approach to the absolute $C_n^2$ calibration, see  \cite{Valenzuela2014}.
\begin{figure}
\hspace{-0.4cm}
\includegraphics[scale=0.48]{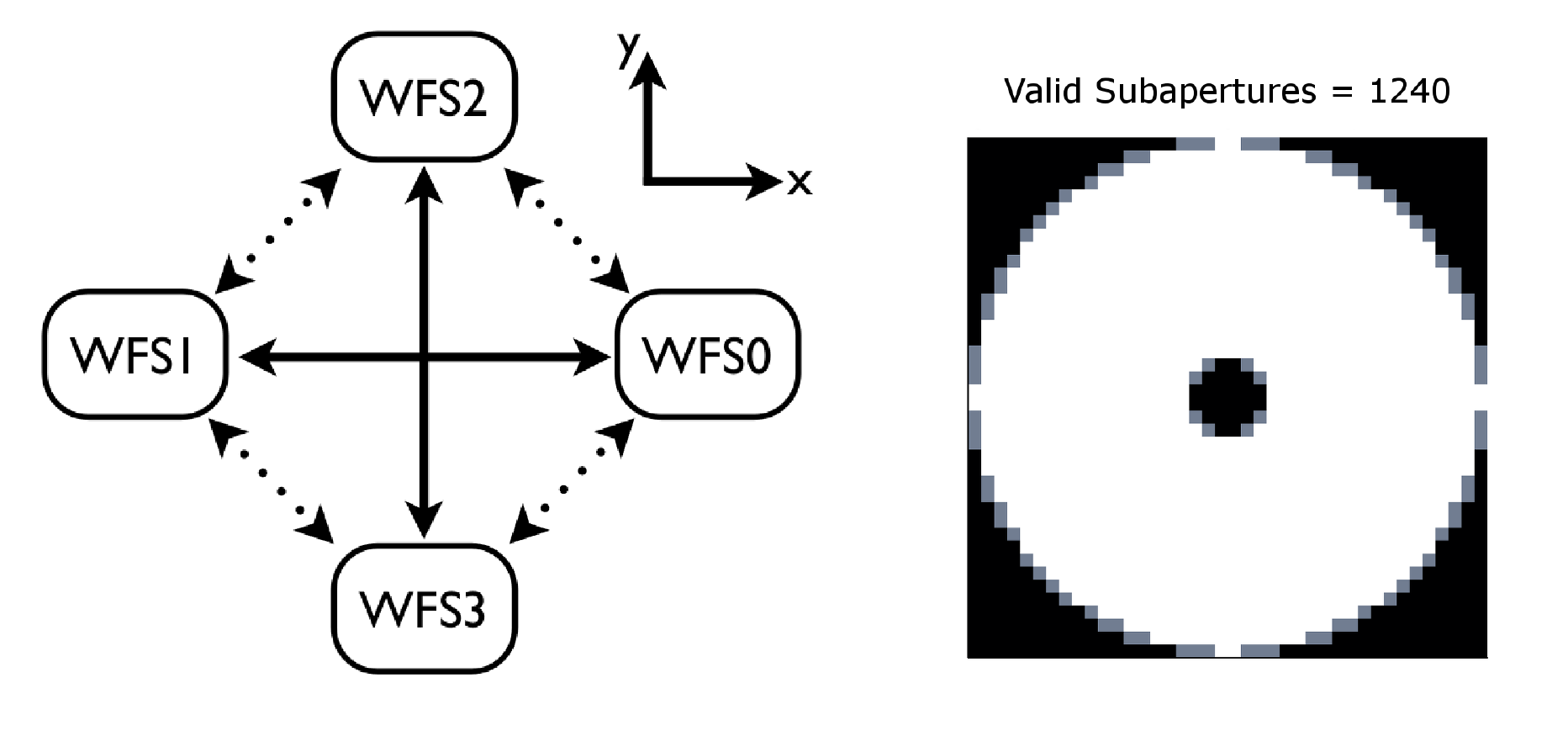}
 \caption{Left: AOF geometry for the four Shack-Hartmann WFSs. Solid and dotted lines indicate the high- and low-resolution baselines, respectively. The reference frame for {\it x} and {\it y} slopes directions is also shown. Right: a picture of one AOF WFS. Fully and partially illuminated SAs are denoted in white and grey, respectively. They altogether define the so-called ``default'' correlation mask, with 1240 SAs.}
 \label{fig:base}
\end{figure}

The AOF geometry is shown in Fig. \ref{fig:base}, along with the directions we call  {\it x}  and {\it y} for the WFSs slopes reference frame. We can define two sets of resolution baselines: one refers to the larger angular separation connecting WFS0-WFS1 and WFS2-WFS3 (shown with solid lines, that we call `high-resolution' or HR baseline), and the other to the smaller angular separation connecting laterally the WFSs (shown with dotted lines, called `low-resolution' or LR baseline).  Fig. \ref{fig:base} (right) shows how the AOF WFSs look like in terms of valid (white+grey regions) and only fully illuminated SAs (white region), defining correlation masks for these configurations (see explanation below).  The grey pixels denote the partially illuminated SAs, with fluxes above 50 per cent of the those reached in fully illuminated SAs.  

\begin{table*}
\hspace{-2.0cm}
 \begin{minipage}{150mm}
  \caption{Maximum altitudes and spatial resolutions probed by the profiler using all 1240 valid SAs (default mask). NFM: Narrow Field Mode; WFM: Wide Field Mode. Columns 3 \& 7: angular separation of baselines. Columns 4 \& 8: number of adopted bins for baselines. Columns 5 \& 9: maximum altitudes for baselines given by the profiler, at $z=30\degree$. Columns 6 \& 10: ranges of spatial resolution in baselines,  at $z=30\degree$.
}
  \label{tab:aofsystems}
  \begin{tabular}{@{}lccccccccc@{}}
  \hline
    &    & \multicolumn{4}{c}{Low Resolution}& \multicolumn{4}{c}{High Resolution}\\
     & Mode   & $\theta_\mathrm{LR}$ & $N_\mathrm{LR}$ & $h_\mathrm{max,LR}^{z=30\degree}$ & $\delta h_\mathrm{LR}^{z=30\degree}$ & $\theta_\mathrm{HR}$ & $N_\mathrm{HR}$  & $h_\mathrm{max,HR}^{z=30\degree}$ & $\delta h_\mathrm{HR}^{z=30\degree}$ \\
   && & & [km] & [km] & && [km] & [km]\\
   GALACSI NFM & LTAO & 14.1" & 27/0$^a$  & 45.5/-$^a$ & 3.4-0.9/-$^a$ & 20.0"   & 38/20$^a$ & 37.9/24.5$^a$ & 1.7-0.6/1.7-0.9$^a$\\
   GALACSI WFM & GLAO & 1.51' & 27 & 12.4 & 0.55-0.41 & 2.13'  & 38 & 9.14 & 0.28-0.22 \\
   GRAAL       & GLAO & 8.20' & 27 & 2.49 & 0.102-0.096 & 11.6' & 38 & 1.76 & 0.051-0.049 \\ 
\hline
\end{tabular}
$^a$The slash symbol separates the maximum possible and the actually adopted number of bins, altitudes or resolution ranges for GALACSI NFM, respectively. In this case, only the HR bins are considered in the analysis. 
\end{minipage}
\end{table*}

The profiler algorithm relies on the cross-correlation between WFSs that we call ``A" and ``B". Its main equation is reproduced below \citep[and references therein]{2014MNRAS.440.1925G}:
\begin{equation}
T^\mathrm{AB}(\Delta u, \Delta v, \Delta t) =  \frac{\left< \sum_{u,v} S_{u,v}^\mathrm{A}(t).S_{u+\Delta u,v+\Delta v}^\mathrm{B}(t+\Delta t)\right>}{O^\mathrm{AB}_{\Delta u, \Delta v}}
\end{equation}
\noindent where $S_{u,v}$ is the $x$ or $y$ slope contained in the SA denoted by the index $(u,v)$ at time $t$, and $\Delta u$ and $\Delta v$ are displacements -- in integer numbers of SAs -- in the WFS grid. In this work, as mentioned previously, the frame delay denoted by $\Delta t$ is always taken as zero. The summation $\sum_{u,v}$ is carried out over all overlapping SAs, and the angle bracket symbols denote the time series average. The definition of the correlation masks is contained in $O^\mathrm{AB}_{\Delta u, \Delta v}$, which compensates for the unequal number of correlated SAs as one moves away from the center of the correlation maps. For the AOF, which suffers no fratricide effect, and in particular for this set of simulations without the presence of spiders, $O^\mathrm{AB}$ is the same for all pairs of WFSs\footnote{This assumption has to be re-evaluated for the ASSIST bench or telescope measurements because of the presence of spiders.}. 

The next step comprises a deconvolution for the impulse response of one single average WFS:
\begin{equation}
T^\mathrm{AB*} = \mathcal{F}^{-1}\left[\mathcal{F}[T^\mathrm{AB}]/\mathcal{F}[(T^\mathrm{AA}+T^\mathrm{BB})\times 0.5]\right]
\end{equation}
\noindent where $\mathcal{F}$ and $\mathcal{F}^{-1}$ denote the Fourier and the inverse Fourier transforms, respectively, and $T^\mathrm{AB*}$ is the deconvolved cross-correlation map.
The profile is extracted from the deconvolved maps along the directions that connect the baselines, starting from the center. The redundant baselines sharing the same separating angles of laser guide stars are combined to provide the HR and LR measured profiles. 

The absolute calibration of the above HR and LR profiles requires reference data, which is accomplished by simulating a large number of independent phase screens with pre-defined turbulent parameters. The WFS slopes derived from such phase screens are combined to emulate a flat profile observed in a large number of frames, from which the average angle-of-arrival (AA) variance is derived. A similar analysis as that carried out for the measured data is performed, obtaining a calibration profile which can be used to normalize the measured one. Finally, multiplying such normalized measured profile by the AA-variance from the calibration data provides the stratified distribution of slope variances in the atmosphere. This is then corrected by the cone effect and converted into a distribution of Fried parameters per altitude bin using a parameterisation of the von K\'arm\'an atmospheric model of the type:
\begin{equation}
\sigma_\mathrm{AA}^2 \propto ~\lambda^2~r_0^{-5/3}~d^{-1/3} \times f(L_0)
\label{eq:r0}
\end{equation}
\noindent in units of rad$^2$, with $d = 0.2$ m (SA size), $\lambda = 589$ nm and $f$ being a function of $L_0$ only. Notice that the calibration procedure makes an assumption on the value of $L_0$ that should reflect the outer scale of the analysed data. The distribution of outer scales measured in Paranal shows a concentration around 22 m \citep{2000A&AS..144...39M}. The AA-variance from a von K\'arm\'an model has a strong dependence on this parameter on smaller scales, compromising the absolute profile calibration if a good estimate of $L_0$ from data is precluded. For this work we used a parameterisation derived by \cite{Conan:00} -- with $f$ presenting a dependence on $L_0$ given as powers of -1/3, -2 and -7/3, and checked against the AA-variances measured from the calibration phase screens -- in contrast to the original work on GeMS, which adopted a Kolmogorov model ($f$=1). 

The $r_0$ distribution obtained from the adopted parameterisation is easily converted to the HR and LR $C_n^2$ profiles of the {\it seen} turbulence, in units of m$^{1/3}$. On the other hand, the integration of these $C_n^2$ profiles and their conversion to $\lambda = 500$ nm can provide the relevant Fried parameter of the integrated {\it seen} turbulence for both resolutions. In order to get rid of any edge effects that occur in the cross-correlation maps, only bins from 1 to $N_{HR}$ and $N_{LR}$ are kept, discarding those of higher altitudes. For the GLAO modes and assuming that the analysis was carried out on all valid SAs (i.e. using the default mask), as default we build a combined profile by keeping the first 38 HR bins and populating the higher altitudes with the remaining selected LR bins. This generates 46 altitude bins in total, from which a combined Fried parameter can also be derived. For the LTAO mode we discard the LR profile -- too poor in resolution -- and keep only the 20 first bins of the HR one. The main parameters for the profiler in each mode are shown in Table \ref{tab:aofsystems}.

\begin{figure}
\hspace{-0.4cm}
\includegraphics[scale=0.31]{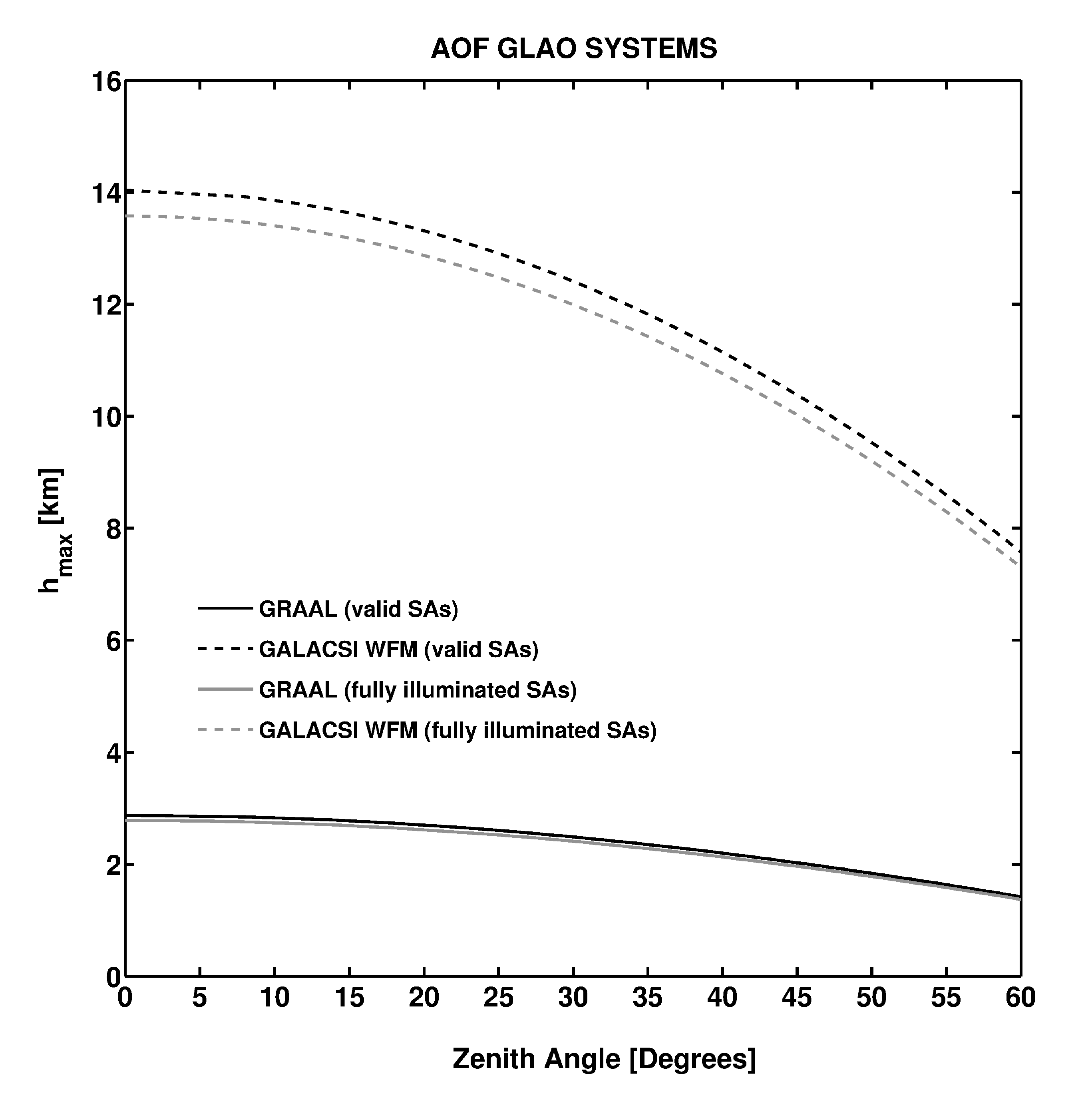}
 \caption{Maximum altitudes probed by the AOF GLAO modes as a function of the zenith angle, assuming that all valid SAs (black) and only the fully illuminated ones (grey) are used for the correlation mask. The AOF LTAO mode is not shown here since its maximum altitude reaches much beyond 20 km  for both mask cases, no matter the zenith angle of the observation ($0\degree \leq z \leq 60\degree$).}
 \label{fig:fig1}
\end{figure}

The altitude bins (in metres) can be calculated from the following formula (for both HR and LR baselines):
\begin{equation}
\label{eq:altitude}
h_m = \frac{m~d~h_\mathrm{Na}}{\sec{z}~h_\mathrm{Na}~\theta' + m~d} + h_0~~~~~m=0,1...N_b-1
\end{equation}  
\noindent where $\sec{z}$ corresponds to the airmass of the observation (the secant of its zenith angle $z$), $h_\mathrm{Na}$ to the altitude of the Sodium layer, $\theta'$ to an angle -- in radians -- related to the baseline under consideration (from Table \ref{tab:aofsystems}, we set $\theta'=\theta_\mathrm{HR}$ and $\theta'=\theta_\mathrm{LR}/\sqrt{2}$ for HR
and LR baselines, respectively). The parameter $d$ (=0.20 m) is again the SA size and $N_b$ the number of sampled bins (40 for HR and 28 for LR). Finally, the parameter $h_0$ is a zero point scale for the binned altitudes. In the case of the AOF, $h_0 = -89$ m accounts for the conjugation of the entrance pupil to this altitude below the ground, which is associated to the DSM location. 

Fig. \ref{fig:fig1} shows $h_\mathrm{max}$, defined as the maximum height achieved by the profile, as a function of the zenith angle, for only the AOF GLAO modes (GRAAL and GALACSI WFM). With the two GALACSI modes we are able to cover -- although less strictly in WFM -- the bulk of the turbulence contained in the troposphere. GRAAL offers extremely good spatial resolutions (as shown by $\delta_\mathrm{HR}$ and $\delta_\mathrm{LR}$ for $z$ = 30$\degree$, see Table \ref{tab:aofsystems}) but fails to provide information for altitudes higher than 2.9 km (or $\sim$1.4 km, in the extreme case of $z$ = 60$\degree$). As mentioned previously, with GALACSI NFM we limit ourselves to 20 HR bins for them being enough to probe, with the best resolution, up to 24.5 km at $z$ = 30$\degree$. This way we avoid degrading our results with the noisier and negligible higher turbulence layers. 

The simulations of all `measurement' data shown in this work have been generated with the Octopus end-to-end simulation tool. This is a parallelized code that runs on a Linux cluster dedicated to simulations of adaptive optics systems at ESO. A good comparison of Octopus with other existing simulation tools in the AO community can be found in \cite{lelouarn2010}. Throughout this work we adopted a pupil size of 400 pixels to represent the 8 m telescope aperture, with a central obscuration of 14.5 per cent. The AOF WFSs contain SAs with 6$\times$6 pixels, with a field of view (FoV) of 0.81''\,pixel$^{-1}$. The Low Light Level CCD detector was simulated by doubling the amount of photon noise and adding 0.2 electrons per pixel of background noise to take into account the readout noise left after the amplification of the detector signal.  For simulations mimicking realistic conditions (section \ref{sec:allmodes}) 10 turbulent layers were adopted, distributed up to 14 km and allowed to displace in random orthogonal directions with user-defined turbulence weights and wind speeds. Turbulent phases have been generated on a support of 8192 pixels length, meaning that they wrap up for wind speeds larger than 11 m\,s$^{-1}$ in typical simulation timespans (15 seconds). However, because such layers are concentrated in higher altitudes and superpose only with smaller weights to the total turbulence, any repetition effects should be negligible. For test cases such as the single layer analysis (see next section) the wrapping of phase screens would, at maximum, set a conservative error to the results.

\section[]{PERFORMANCE ASSESSMENT CASES}
\label{sec:tests}
GALACSI WFM, the intermediate spatial resolution case, is the mode under inspection in this section. Other AOF modes are treated later, once we have an insight on the limitations of the profiler in this intermediate case. In order to disentangle the effects introduced by the POL computation from GLAO/LTAO CL modes and the outcome from the profiler, we opted for analyzing in this section only OL simulations. Taking this approach, we bear in mind that we may incur some non-linear behaviour of the WFSs. All simulations in this section have been run in the WFS high photon flux regime. Turbulence parameters of ($r_0, L_0$) = (0.10, 25) m have been used. We also assumed $z$ = 0$\degree$ so the probed altitudes were at their maximum, and $h_0$ = 0 m. 

\subsection[]{Response to Different Turbulence Distributions}
\begin{figure*}
\includegraphics[scale=0.50]{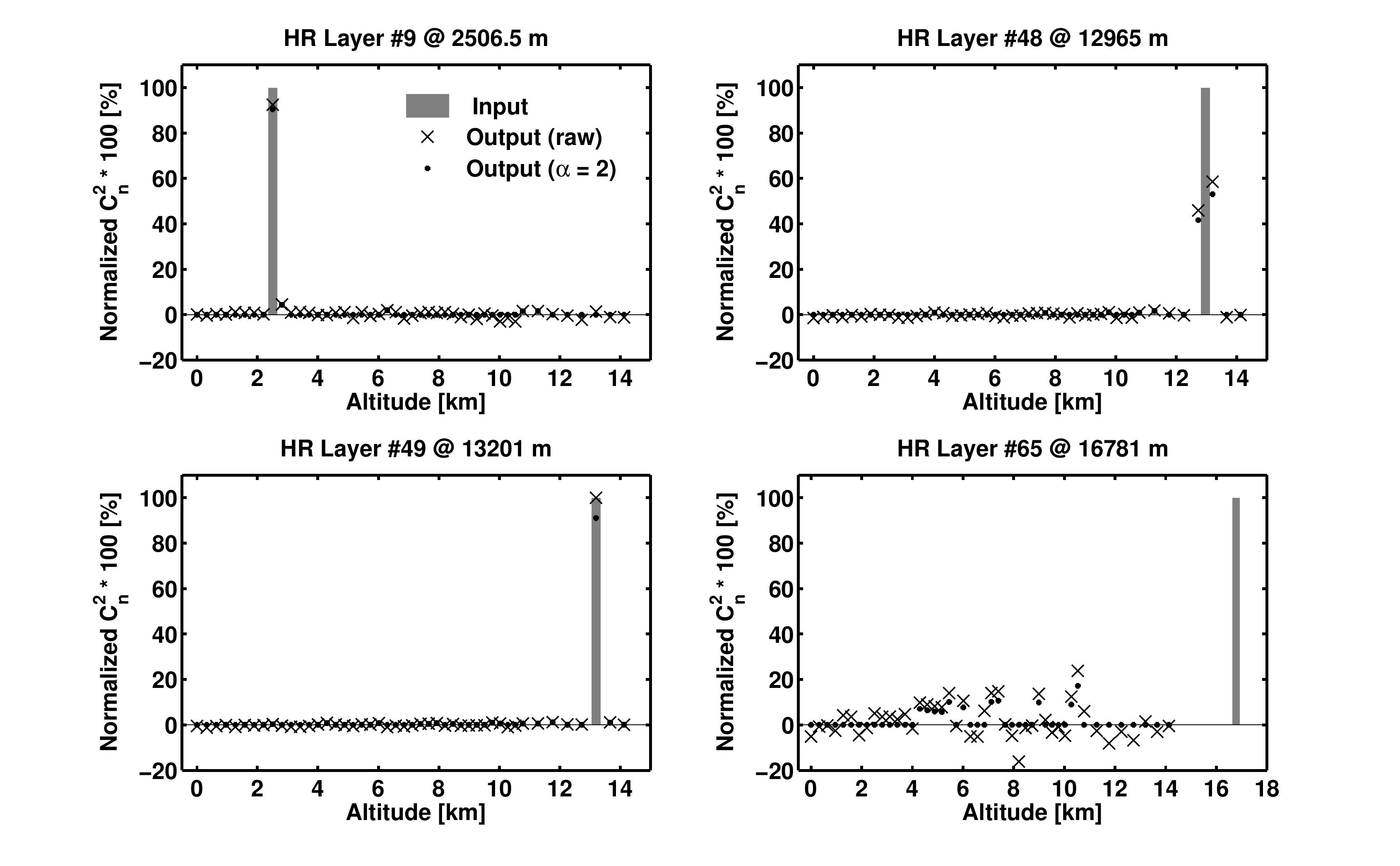}
 \caption{Examples of the results from the $C_n^2$ profiler for simulations of single layers at different altitudes. The grey bar represents the input value (=100 per cent in one layer) whereas the crosses and dots show the output results as given by the profiler and noise-filtered, respectively. Layer \#9: exact HR bin; layer \#48: falling between two high altitude LR sampled bins; layer \#49: falling at an exact sampled high altitude LR bin; layer \#65: altitude not sampled by the profiler. The apparent high levels of noise, produced by a direct normalization of the profile, can be corrected by taking into account the unseen turbulence (see later).}
\label{fig:onelayer}
\end{figure*}
\vspace{0.5cm}
\subsubsection{Single-Layer Profiles}

For this section, simulations comprised of profiles of one single atmospheric layer concentrating all turbulence have been analysed. The layers were simulated at the exact altitudes of the GALACSI WFM HR bins as given by equation (\ref{eq:altitude}) with $\theta'=\theta_\mathrm{HR}$. Notice that turbulent layers beyond the altitude corresponding to $N_b=56$ are not supposed to be probed by the profiler. Fig. \ref{fig:onelayer} shows typical examples of the profiler response when having the turbulent layer located at the 9$^\mathrm{th}$,  48$^\mathrm{th}$,  49$^\mathrm{th}$ and 65$^\mathrm{th}$ HR altitude bins, with the integrated $C_n^2$ output from the profiler normalized to 1. Input layers are given by the grey bars and the raw output from the profiler by the crosses. In terms of qualitative response, the profiler succeeds in reconstructing the input layers at their locations. We notice different types of response: 
\begin{enumerate}
\item Strong response in one output layer when the input layer falls exactly in a HR or a LR profile bin (e.g. layers \#9 and \#49, respectively).
\item Strong response in 2 layers when the input layer falls in a HR-like bin not sampled by the HR part of the profiler but it is still within the range of the combined profile (e.g. layer \#48). In other words, the sum of adjacent bins approximately matches that of the input profile.
\item No response when the input layer is beyond the reconstructed profile (e.g. layer \#65). In this case the normalization of the raw output profile to 1 explains the apparent high level of noise seen in the figure. The noise, in absolute $C_n^2$ value, is actually the same as in the other types of response. Once the {\it unseen} turbulence is accounted for (see discussion later), the apparent noise in the normalized profile drops significantly, looking the same as in the other cases.
\end{enumerate}
Noise is observed in all layers where no turbulence is expected, resulting in an output $C_n^2$ value at the location of the peak with less than 100 per cent.
By selecting a certain number of bin data points that would exclude the turbulent ones (for instance, the $N \sim N_\mathrm{t} - 10$ smallest values of the output profile, where $N_\mathrm{t}$ is the total number of combined bins probed by the profiler), we can compute their median $M_P=\mathrm{median}(P)$ and median deviation values defined by
\begin{equation}
\sigma_P = \sqrt{\mathrm{median}(P-M_P)^2}
\end{equation}
and devise a filter that sets to zero all $C_n^2$ values lower than $M_P+\alpha.\sigma_P$. The parameter $\alpha$ is a threshold to be defined and that depends on the amount of noise to be filtered. The purpose of this procedure is to remove unrealistic profile values e.g. those with negative intensities. Once the filtering is applied, the $C_n^2$ should be renormalized to 1.
Applying the filter (with $\alpha=1$) to every profile obtained from simulated data with one layer at a visible altitude
($\la$14 km), we obtain median values of the noise which are very small (0.13 per cent on average) and oscillating close to 0 per cent. The median deviation of the noise varies
between 0.4 and 0.6 per cent with an average value of about 0.5 per cent. This value can be
viewed as the noise `standard deviation' of the $C_n^2$ profile estimation
in layers without turbulence. On the other hand, the ratio of the output turbulence in a determined layer to the total turbulence -- estimated from
a simulation with 100 per cent input turbulence exactly at this bin (or in the 2 adjacent layers when in
LR) -- lies between 75-90 per cent for the HR bins and around 100 per cent for the LR bins. In other words, the peaks are detected in all resolutions but their strength is lower than the input one, specially in HR. A correction with a $\alpha=1$ filter can make the results worse. This
shows that by not filtering enough the noise the whole profile is biased towards positive values,
decreasing the relative amplitude of the peak; for a noise filter with $\alpha=2$ the results turned out to be similar to the non-filtered ones.

Bear in mind that the layers above $\sim$14 km are not sampled.
One should filter even more the noise or broaden the bins but that may not be a valid
solution when using real data for losing resolution. The better response of the higher layers suggests that the
LR profile is less noisy and gives more accurate response to the input turbulence. A suggestion 
would be that, when $C_n^2$ accuracy is more important than spatial resolution, one should
use the LR profile only; this could be the case, for instance, of GRAAL profiles.

If the turbulence is distributed above and below the highest probed altitude $h_\mathrm{max}$, the Fried parameter $r_0^\mathrm{cn2}$, defined as the one measured by the profiler, is overestimated. On the other hand, $r_0^\mathrm{wfs}$ estimates made from the WFSs AA-variances -- properly corrected by the cone effect -- should reflect the total turbulence along their lines-of-sight, being closer (ideally equal) to the input value of the simulations\footnote{This still might not be true because of the limited time sampling and linearity/spot truncation effects of the centroid algorithms.}. Equation (\ref{eq:r0}) for a von K\'arm\'an model implies that the computation of these parameters depend on an assumption for the $L_0$ value of the input data, a tricky parameter to both simulate and estimate with the current algorithm. This has an impact on the absolute calibration of the profile but fortunately not on its shape. For that reason, and because most applications of the profiler to optimize the AOF performance only require information about the relative $C_n^2$ distribution, {\it in this work we restrict ourselves only to the analysis of the normalized turbulence profiles}. This means that $r_0^\mathrm{wfs}$ and $r_0^\mathrm{cn2}$ in this work are just used for a correction of the unseen turbulence. We will return to this discussion when analyzing multi-layer profiles.

\vspace{0.5cm}
\subsubsection{Multilayer Profiles}

\begin{figure}
\includegraphics[scale=0.58]{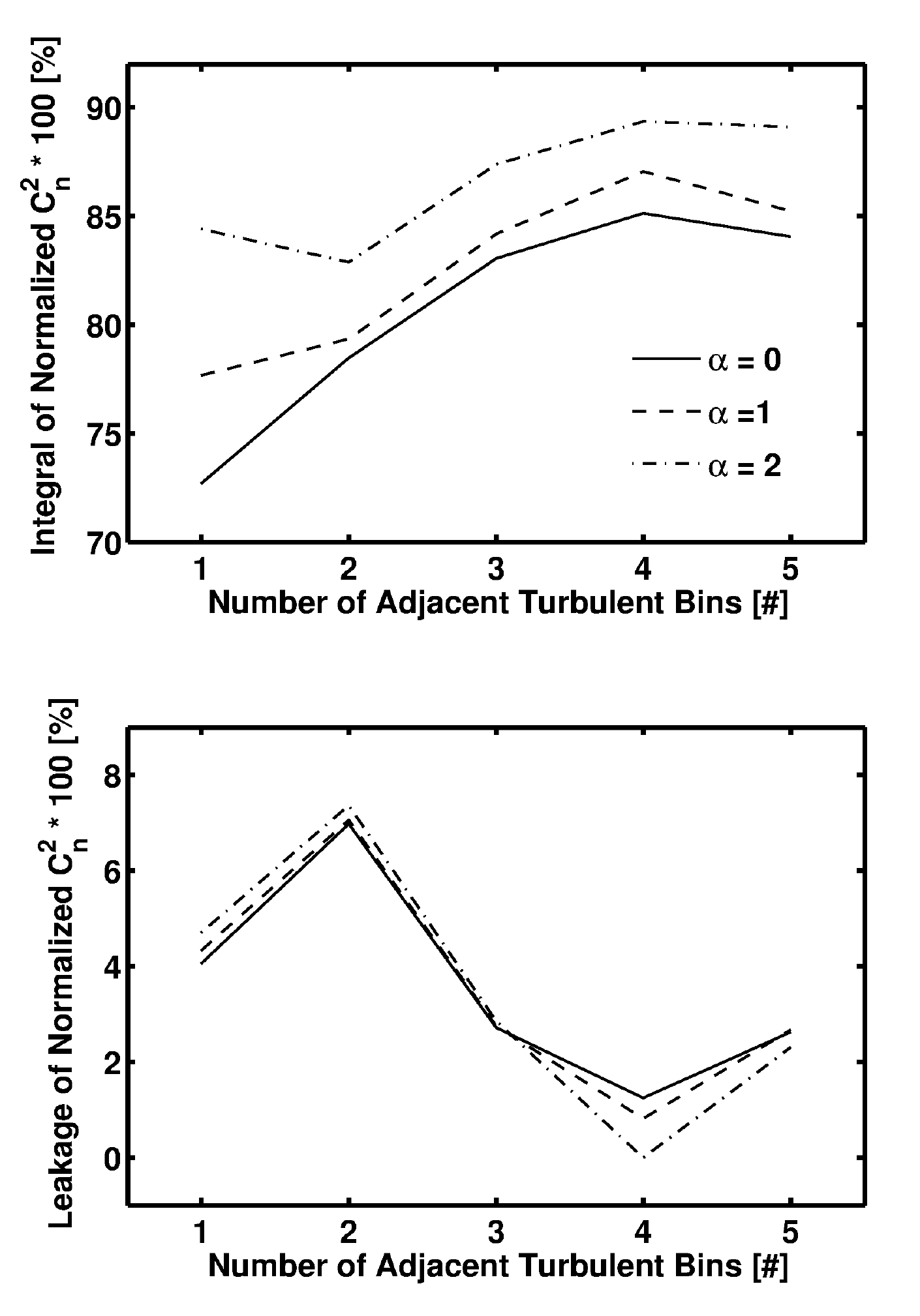}
 \caption{Top: integral of the normalized $C_n^2$ profile, only in the turbulent adjacent bins. Bottom: integral of the normalized $C_n^2$ profile, in the vicinity of the turbulent bins. Curves show the results for an applied noise filter of $\alpha =$ 0, 1 and 2. The results are for the HR part of the profile.}
\label{fig:adjacentbins}
\end{figure}

A natural extension to the single layer analysis is the one of a few adjacent layers. This is useful to gain an insight on the cross-talk between layers in the profiler response. There is some evidence of this behaviour in Fig. \ref{fig:onelayer}, with turbulence input in bin \#9, where we see a small `leakage' of the layer strength to the neighbouring higher altitude bin. Input profiles composed of 1, 2, 3, 4 or 5 layers (containing each an equal share of the total turbulence) located at the exact altitudes of an output HR profile bin close to 5 km and the bins immediately above have been simulated to further check this behaviour. With $\alpha=0$ for the noise filtering, the integrated value of the output turbulence in those layers tends to increase from 73 to a maximum of $\sim$84-85 per cent with the increase in the number of turbulent bins from 1 to 5 (Fig. \ref{fig:adjacentbins}). The results also show some leakage of intensity to neighbouring layers that can reach $\sim$7 per cent, tending to decrease for a larger number of input turbulent bins. Using $\alpha=1$ and 2 for the noise filtering produces slightly better results, although with final integrated intensities still below 90 per cent. In summary, this test has shown that the profiler seems to respond better to a turbulence which is not concentrated in isolated thin slabs, at least in the HR part of the profile (for LR the convergence is better, as mentioned earlier in the single-layer case).
\begin{figure}
\includegraphics[scale=0.585]{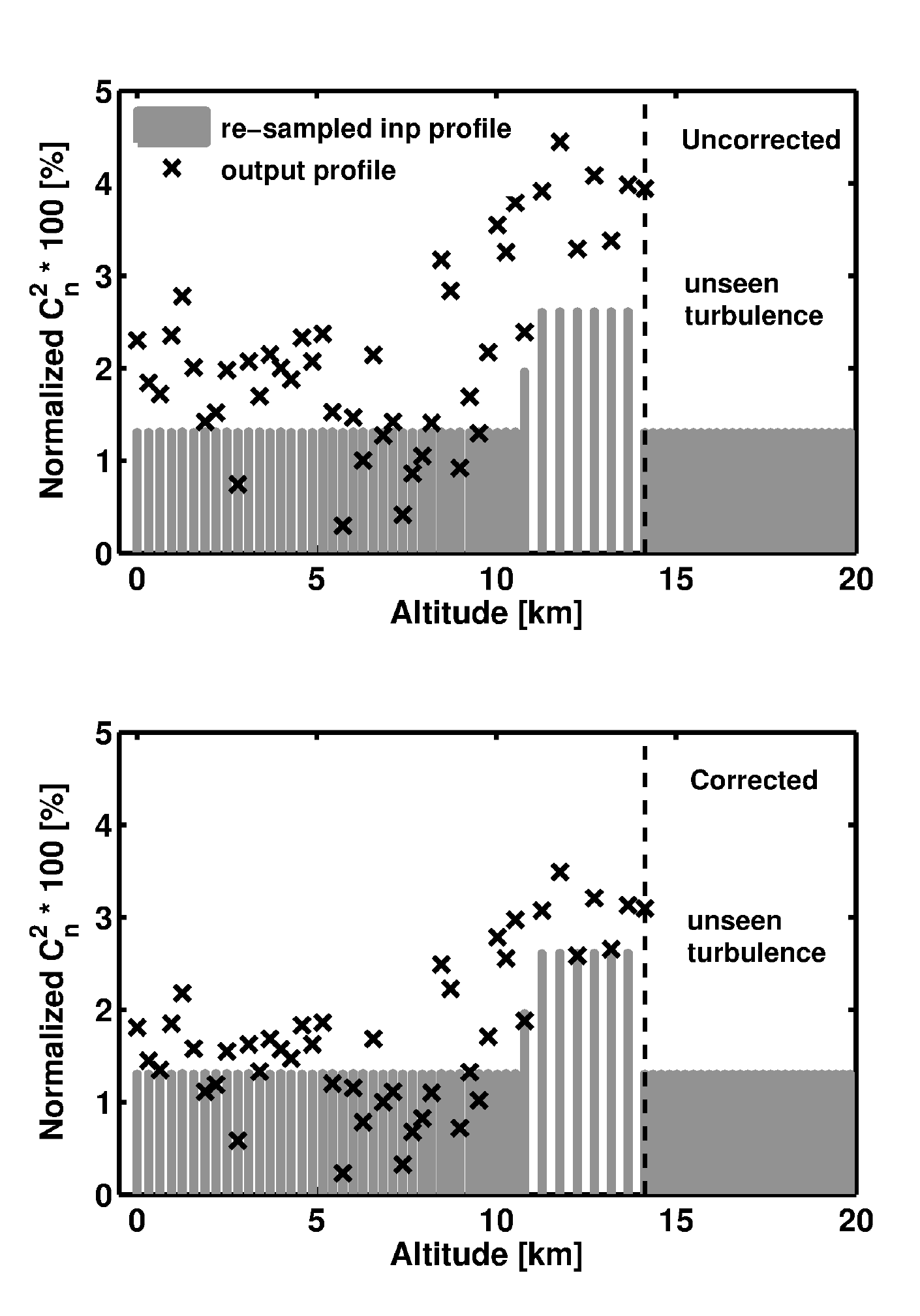}
\vspace{-0.8cm}
 \caption{Response of the profiler to a flat $C_n^2$ distribution. Top: uncorrected by unseen turbulence; bottom: corrected by unseen turbulence, using the prior knowledge of the $C_n^2$ distribution from the simulations. The rms between the output profiles minus the resampled input ones is 1.12 and 0.7 per cent for top and bottom cases, respectively. }
\label{fig:flatprof}
\end{figure}

A case of a flat input profile composed of $N_b = 80$ layers has also been analysed, with each layer located at the exact altitudes of the HR profile bins. In practical terms, this means that there is a significant turbulence fraction which is not seen by the profiler.
A first inspection shows that the average level of the estimation turns out to be good, but the output profile seems to increase between $\sim$10-14 km. This happens because at those altitudes each output bin (which is LR) covers 2 layers of the input profile thus doubling their total turbulent strength. A much fairer representation consists in resampling the simulation layers intensities to the output bins, assuming a linear interpolation based on results from tests with sub-bin displacements. To do so, each simulated layer intensity is spread into the 2 nearest output bins $i$ and $i+1$ with a ratio proportional to the distance of $h_{i+1}$ and $h_i$ to that layer altitude, respectively. 
For instance, if $H_k$ (= input layer altitude) lies between $h_1$ and $h_2$ (i.e. $h_1\leq H_k\leq h_2$), the intensity of the input bin will be shared between the two output bins with weights of $(h_2 - H_k)/(h_2-h_1)$ and $(H_k-h_1)/(h_2-h_1)$, respectively. In this representation, input contributions which lie above the maximum probed altitude remain unaffected. Fig. \ref{fig:flatprof} (top) shows the resampled input and the output profiles. The rms of their difference in the sampled bins is of 1.12 per cent (almost the intensity of the input signal in each of the 80 HR bins, i.e. of 100 / 80 $\sim$ 1.30 per cent).

A still better match of the output profile can be obtained if we compensate the output profile by the unseen turbulence. In this specific case, the Fried parameter input in the Octopus simulations, called hereafter $r_0^\mathrm{sim}$, is 10.3 cm. On the other hand, $r_0^\mathrm{cn2}$ is 14.6 cm and the average Fried parameter obtained from the WFSs AA-variances, still uncorrected by the cone effect, is 12.5 cm. An equivalent turbulence altitude can be estimated using the formula:
\begin{equation}
\label{eq:h}
\overline{h} = \left(\sum{C_n^2(h_i) h_i^{5/3}} / \sum{C_n^2(h_i)}\right)^{3/5}
\end{equation}
\noindent where the input $C_n^2$ distribution can come either from the profiler output or from {\it a priori} knowledge of the input $C_n^2$ distribution (e.g. from simulations). The cone effect factor can then be approximated by $f_c \sim 1-\sec{z}~\overline{h}/h_\mathrm{Na}$.
Such a factor produces a $r_0^\mathrm{wfs}$ estimate of 11.3 and 10.9 cm, if we consider $\overline{h}$ equal to 8.47 km (using all 46 output profile bins and intensities) and 11.8 km (using all 80 seen+unseen simulated bins and intensities) for such a correction, respectively. In the case of real measurements, the output profile should thus be multiplied by $(r_0^\mathrm{cn2}/r_0^\mathrm{wfs})^{-5/6}$ to account for the unseen turbulence. Notice at this point that any dependence on $L_0$ is cancelled out in a normalized profile. Fig. \ref{fig:flatprof} (bottom) shows the improvement obtained in the output profile if we apply the correction considering all simulated bins. The rms of the residuals after the cone effect and unseen turbulence correction drops to 0.7 per cent in this case. Naturally the improvement can be even better if one provides an independent and better estimate of the total $r_0$, since the normalization factor of $(14.6/10.3)^{-5/6}$ = 0.75 leads to a better correction of the unseen turbulence (rms of the residuals $\sim$ 0.5 per cent).
In summary, even not constituting a realistic scenario because of its profile flatness, this case is helpful to validate the bin resampling procedure and the correction for the unseen turbulence. It confirms very clearly the 0.5 per cent figure for the noise standard deviation of the profile estimation in this extreme case of large equivalent altitude turbulence. It is also worth pointing out that the only AOF mode likely to be significantly affected by unseen turbulence is GRAAL because of its low altitude coverage (maximum height of 2.9 km at zenith).

\begin{figure}
\vspace{0.4cm}
\includegraphics[scale=0.58]{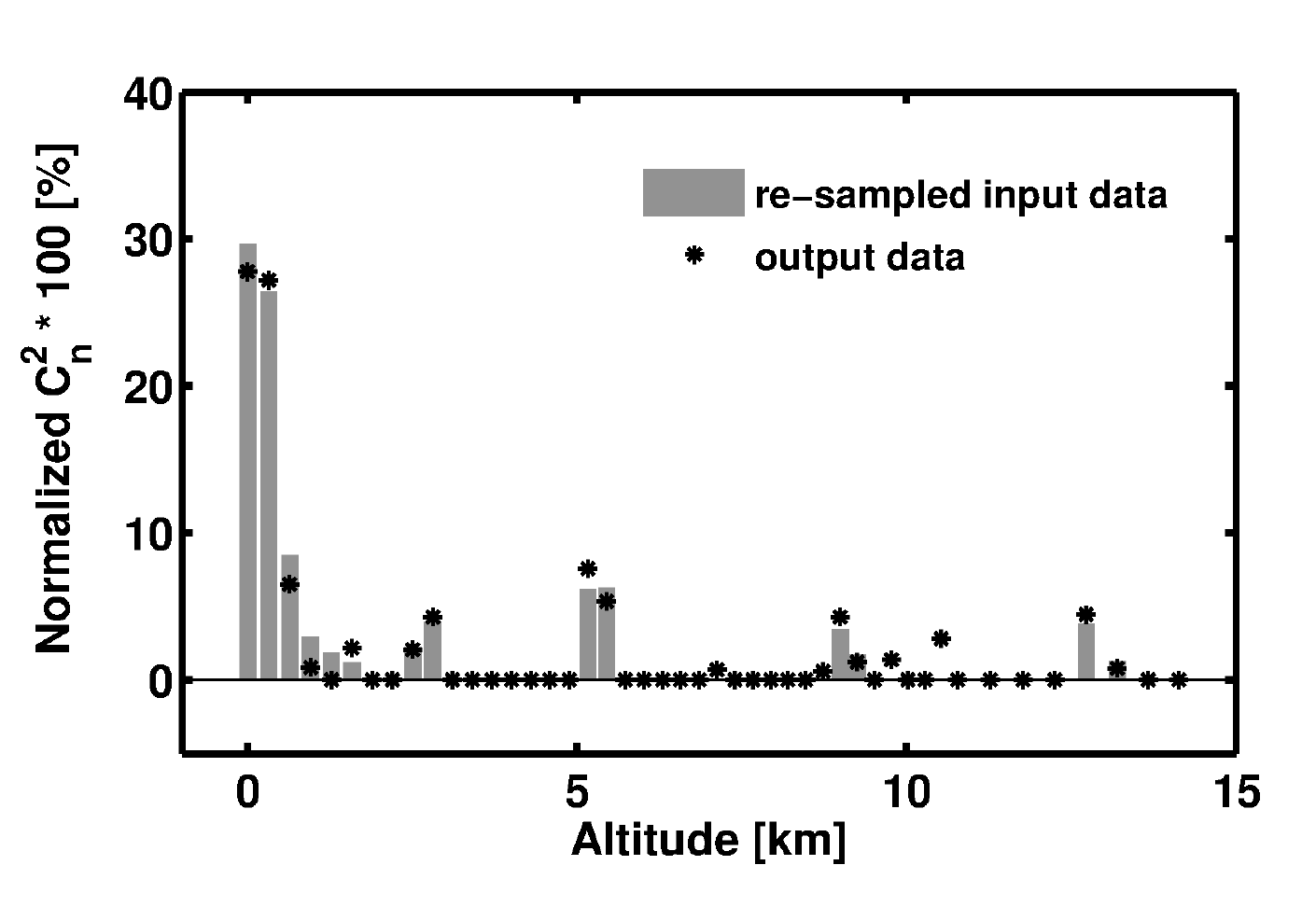}
 \caption{Hypothetical profile used in the analysis from sections \ref{sec:undersampling} to \ref{sec:Na}, resampled to the output bins of the profiler. The output profile shown here refers to the profiler results for a GALACSI WFM OL simulation of high flux (800 photons SA$^{-1}$ frame$^{-1}$), an undersampling of $\tau=$10 and a dataset length of 15000 frames. A noise filtering of $\alpha = 1$ was applied.}
\label{fig:hypoprof}
\end{figure}

\subsection[]{Dependence on Temporal Data Sampling}
\label{sec:undersampling}

\begin{figure}
\includegraphics[scale=0.58]{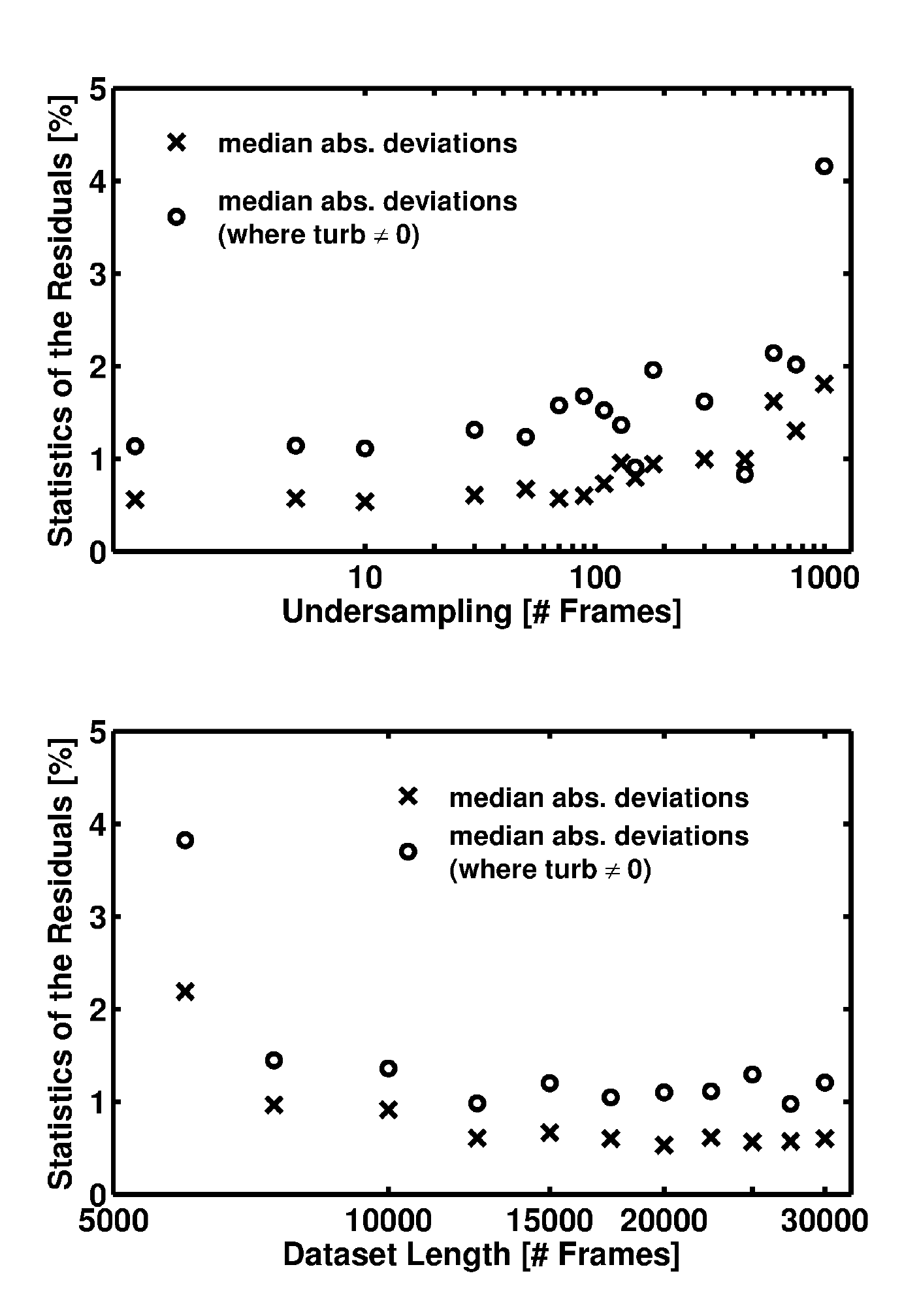}
 \caption{Statistics of the residuals: median and maximum absolute deviations for different samplings. Top:  taking only a certain percentage of frames in the cross-correlation computation; bottom: taking different dataset lengths, for a fixed percentage of frames ($\tau = 10$). }
\label{fig:subsampling}
\end{figure}
The quantities used in the derivation of the $C_n^2$ profiles are temporal statistical averages.  Therefore, it may be possible to achieve in the post-processing stage the same results using only a smaller subset of the original number of frames. 

A hypothetical profile as shown in Fig. \ref{fig:hypoprof} is adopted for this test. As a quality criterium we use the median absolute deviations of the residuals (resampled normalized input profile minus the normalized output profile).  Since this hypothetical profile has no turbulence contribution beyond 14 km, no unseen turbulence correction is needed, and therefore the integrated turbulence in the probed range is 100 per cent. For the case shown in the figure, considering only bins with non-zero input turbulence, the adopted residual criterium gives 0.9 per cent.

Fig. \ref{fig:subsampling} (top) shows our statistical criterium as a function of the adopted undersampling interval for a dataset length of 15000 frames (15 seconds @ 1 kHz). An undersampling of $\tau$ means that only $1/\tau$ of the total number of frames are taken into account in the cross-correlation computation. It can be clearly seen that for up to an undersampling interval of about 10-50 frames the residuals do not change much but start to do significantly for intervals higher than about 100 frames. The undersampling parameter has an impact on the time required to compute the profile and this can be of great value for the real time optimization of the AO correction.

Another possible test refers to the dataset length. This aspect has an impact on the minimum time required for the acquisition of telemetry data in order to retrieve a reliable atmospheric profile estimate. Fig. \ref{fig:subsampling} (bottom) shows the results for this test, with dataset lengths ranging from 6000 to 30000 frames and adopting a fixed undersampling interval of 10 frames. Apparently the dataset length starts to affect the output when the number of acquired frames is less than $\sim$10000 (@ 1 kHz loop rate). For comparison, the GeMS circular buffer files analysed in \cite{2012MNRAS.427.2089C} had a maximum size capacity of 24k frames which, in a loop rate tuned to compensate for the photon return ($<$800 Hz), produced datasets of the order of a few minutes. Regarding the CANARY bench working in MOAO, \cite{morris_ao4elt3} reported an acquisition of 2-3 minutes for the determination of the profile from WFSs telemetry data.
Being conservative, we adopt 15 seconds as our minimum individual `acquisition' time at the nominal loop rate. In real observations, a procedure of averaging profiles obtained from different datasets spanning a couple of minutes might be adopted as well in order to ensure a better statistical result.

\subsection[]{Dependence on Input $L_0$}
\label{sec:L0}
The Octopus simulation studied in this case is the same as the one in the previous section. The analysis, however, differs in the assumption of the $L_0$ for the computation of the $C_n^2$ profile, ranging from 5 to 100 m. 
The normalized output profile is not sensitive to the assumed calibration outer scale, as one could expect, being consistent with median absolute deviations in the residuals of 0.5 per cent. The sensitivity to this parameter is pronounced in the estimation of $r_0^\mathrm{wfs}$ and $r_0^\mathrm{cn2}$, reaching deviations of up to $\sim$35 per cent when the assumed $L_0$ is very different from the effective $L_0$ of the simulations. This confirms that this profiler method alone should not be used to estimate the Fried parameter.  An absolute calibration of the output profile through this method would then require an independent estimate of turbulence parameters. Notice, however, that a newer and smart approach to calibrate the absolute $C_n^2$ profile has been addressed in \cite{Valenzuela2014}; it is based on fitting the covariances profiles, being independent of prior model assumptions. 

\subsection[]{Dependence on Input Na Layer Altitude}
\label{sec:Na}
On the real system, the RTC provides information on the average Sodium layer altitude through the trombone and focus compensator loop.  As an input to the profiler, the same information impacts on the definition of the altitude bins through equation (\ref{eq:altitude}). It has been verified -- with the same simulation data of Fig. \ref{fig:hypoprof} -- that 
when an error is made on the assumed altitude of the Na layer, the positions of the $C_n^2$ layers are mistaken by a small fraction of that amount and the integrated profile amplitude changes slightly. For example, an error of 10 per cent on the input Na layer altitude leads to an error of 1.4 per cent on the $C_n^2$ layers position and to a maximum absolute deviation of 5 per cent from the expected amplitudes. This also implies that the estimation of $r_0^\mathrm{cn2}$ and $r_0^\mathrm{wfs}$ are little affected as well, because their dependencies on the integrated profile and on the cone effect correction variations are negligible. In summary, the $C_n^2$ profiler does not require a good accuracy on the Na layer distance; for this estimation one could assume a median altitude of 90 km and the actual zenith angle, with no need of information from the actual position of the trombones or the focus compensators in the AO modules. Notice, however, that this information does matter for the optimal system correction. Any uncertainties in the Na layer altitude during operations can indirectly affect the $C_n^2$ estimation because of the loss of focus in the WFS cameras and its consequent implications for the slopes estimation (introduction of noise, truncation effects, etc).

\section[]{PERFORMANCE UNDER TYPICAL PARANAL CONDITIONS}
\label{sec:allmodes}

\begin{figure}
\hspace{-0.4cm}
\includegraphics[scale=0.37]{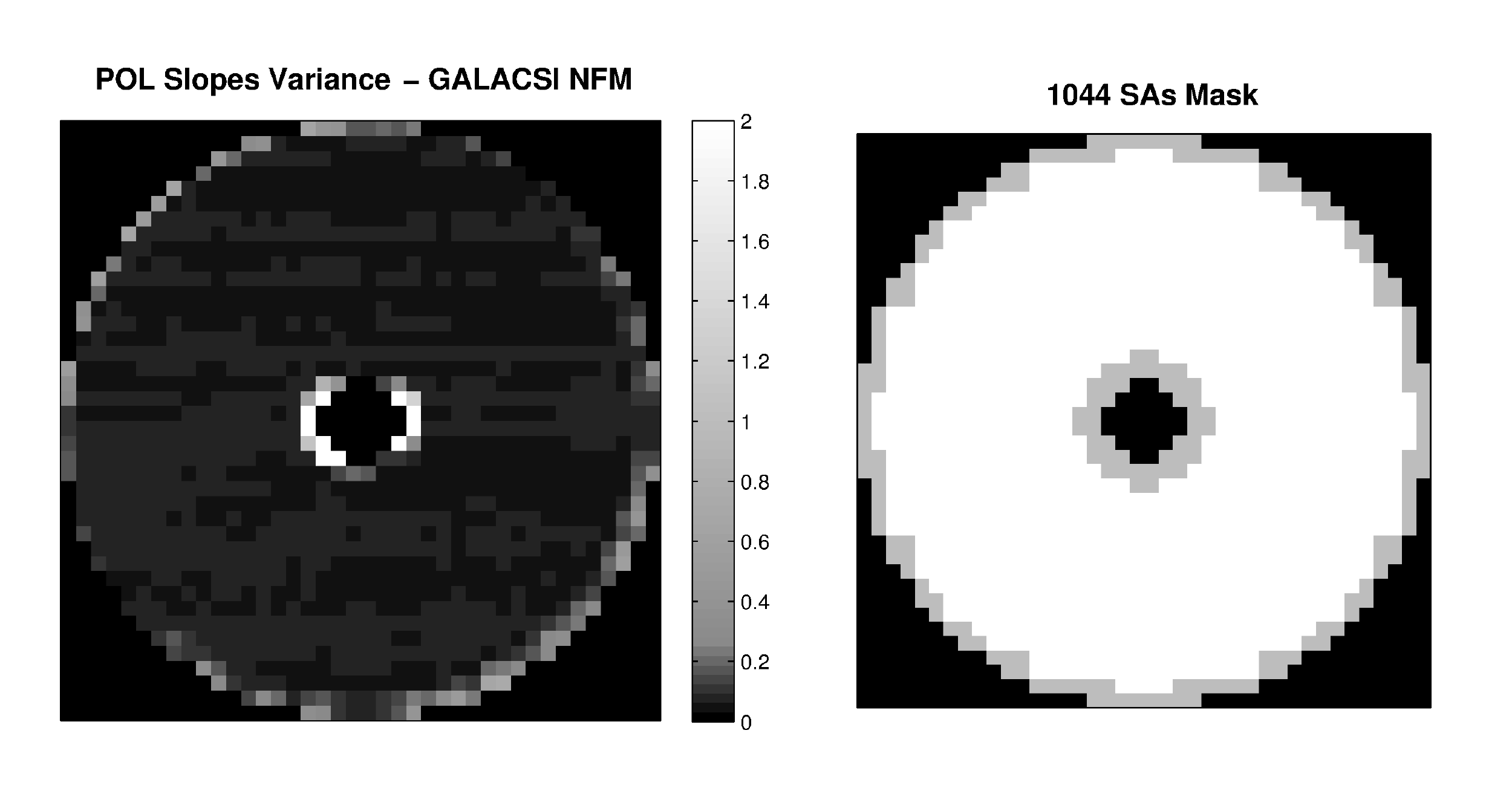}
 \caption{Left: Example of POL slopes variances -- in pixel$^2$ units -- measured in a WFS from simulations of the GALACSI NFM, under a typical Paranal turbulence of 0.6" seeing.  Right: in white, the mask designed to cut off problematic correction regions, containing 1044 SAs to be used in the profiler.}
\label{fig:bigmask}
\end{figure}

Most of the tests shown in previous sections have been carried out using GALACSI WFM in OL, simulated at zero zenith angle and taking into account the conjugation of the telescope pupil to the ground. Hereafter more realistic assumptions are adopted, evaluating CL performances under typical telescope setup and observatory turbulence conditions. The actual value for the altitude conjugation of the DSM is then set to $h_0$ = -89 m. The flux on the WFSs is taken as the nominal one for the AOF, of 80 photons \,SA$^{-1}$\,frame$^{-1}$. This is a safe photon flux regime for robust profiler results as shown by previous tests \citep{AGR2014}. The zonal interaction matrix used in equation (\ref{eq:POL}) for computing the POL slopes is calculated through Octopus in a SCAO configuration and replicated to account for all 4 WFSs when transforming commands into slopes. In GALACSI NFM, ground and `virtual' commands provided by the Octopus LTAO simulations are summed to give one single set of commands to be multiplied by $\mathsf{\mathbf{M}}$. It is important to point out that the reconstructor and the loop gain have yet to be fine-tuned for the maximum Strehl, but stability in the loop was ensured. 

Fig. \ref{fig:bigmask} shows the variance of POL slopes mapped to one of the WFSs, as seen in a CL LTAO case for a typical Paranal profile with seeing 0.6" (@ 0.5~$\mu$m). The figure shows an increased variability in the WFS edges, reflecting a larger uncertainty of the correction in those regions, partially because of discontinued actuators dynamics at the borders. Notice that a wavefront aberration improperly corrected by the DSM at the pupil edges leaves a footprint common to all sensing directions, and this feature translates into a strong cross-correlation when there is an overlap of WFSs maps (i.e. for the lowest altitude bins). Some preliminary tests have shown that even the fully illuminated SAs mask is not effective in dealing with such high variance regions. Therefore, in order to circumvent these spurious effects a new mask based on the variance of the slopes -- see also \cite{2012MNRAS.427.2089C} -- is adopted, with 1044 `good' SAs (Fig. \ref{fig:bigmask}, right). In practical terms, masking out the problematic SAs restricts the altitudes probed by the profiler, shortening them by about 1 bin in maximum height (LR high end of the profile). Applying this mask to GALACSI NFM POL slopes has no limiting effect on its output HR profiles because of the large altitude range achieved in this mode. For GLAO modes the maximum height loss is of about 100 m and 400 m for GRAAL and GALACSI WFM, respectively.

\begin{table}
 \centering
\begin{minipage}{85mm}
  \caption{List of typical turbulence profiles compiled in Paranal. ~
   $\beta$: scaling factor for the layer speeds; ~
   Prob: probability of occurring such conditions in Paranal; ~
   T.Q.: turbulence quality. ~ Notes:  $r_0$ @ 0.5 $\mu$m;  $\tau_0$ @ $z$ = 30$\degree$.
   }
   \label{tab:cn2prof1}
  \begin{tabular}{@{}lccccccl@{}}
  \hline
ID & $S$ & $r_0$  & $\tau_0$  & $\beta$ & $\overline{h}$ &  Prob. & T.Q.\\
        & [~'']   & [m]            & [ms]                  &          & [km]   & [\%] & \\
P01 & 0.4 & 0.186 & 4.6 & 1.29 &  3.73 & 7.0 & median\\
P02 & 0.6 & 0.136 & 3.9 & 1.44 &  2.39 & 6.0 & good\\
P03 & 0.6 & 0.136 & 3.8 & 1.14 &  3.73 & 12.0 & median\\
P04 & 0.6 & 0.136 & 3.9 & 0.92 &  4.88 & 6.0 & bad \\
P05 & 0.8 & 0.116 & 3.0 & 1.44 &  2.66 & 6.5 & good\\
P06 & 0.8 & 0.116 & 3.0 & 1.15 &  3.82 & 13.0 & median \\
P07 & 0.8 & 0.116 & 3.1 & 1.00 &  4.45 & 6.5 & bad \\
P08 & 1.0 & 0.101 & 2.4 & 1.44 &  3.05 & 4.5 & good\\
P09 & 1.0 & 0.101 & 2.5 & 1.19 &  3.81 & 9.0 & median \\
P10 & 1.0 & 0.101 & 2.4 & 1.12 &  4.47 & 4.5 & bad \\
P11 & 1.2 & 0.089 & 2.0 & 1.55 &  2.73 &  3.0 & good\\
P12 & 1.2 & 0.089 & 2.1 & 1.27 &  3.53 & 6.0 & median \\
P13 & 1.2 & 0.089 & 2.0 & 1.36 &  3.69 & 3.0 & bad \\
P14 & 1.4 & 0.074 & 1.4 & 1.64 &  3.53 & 13.0 & median \\
 
\hline
\end{tabular}
\end{minipage}
\end{table}

\begin{table*}
 \centering
 \begin{minipage}{150mm}
  \caption{Typical Paranal turbulence profiles used in the simulations. Pn: profile identification, as in table \ref{tab:cn2prof1}.}
  \label{tab:cn2prof2}
  \begin{tabular}{@{}lccccccccccccccc@{}}
  \hline
Altitude & $v_\mathrm{ref}$ & P01 & P02 & P03 & P04 & P05 & P06 & P07 & P08 & P09 & P10 & P11 & P12 & P13 & P14 \\
 m   & [m\,s$^{-1}$]  & [\%]& [\%]  & [\%]   & [\%] & [\%]& [\%]  & [\%]   & [\%] & [\%]& [\%]  & [\%]   & [\%] & [\%]& [\%]  \\
30    & 5.7 & 70 & 83 & 70 & 53 & 77 & 59 & 41 & 65 & 45 & 16 & 46 & 26 & 0 & 26 \\
140   & 5.1 & 1 &   1 &  1 &  1 & 2  &  2 &  1 &  5 &  4 &  2 & 10 &  8 & 3 & 8  \\
281   & 4.4 & 3 &   3 &  3 & 2  & 4  &  4 &  4 &  7 &  7 &  5 & 12 & 11 & 6 & 11 \\
562   & 3.9 & 5 &   5 &  5 &  5 & 5 &   6 &  8 &  7 &  9 & 12 & 15 & 16 & 23 & 16 \\
1125  & 4.4 & 0 &   0 &  0 &  0 & 0 &   1 &  1 &  0 &  3 & 11 &  1 &  6 & 28 & 6 \\
2250  & 7.2 & 2 &   0 &  2 &  7 & 1 &   5 & 13 &  1 &  6 & 21 &  1 & 10 & 14 & 10 \\
4500  & 14.2 & 5 &   1 &  5 & 11 & 3 &   9 & 15 &  4 & 12 & 16 &  6 & 10 & 14 & 10 \\
7750  & 26.3 & 4 &   2 &  4 &  6 & 2 &   4 &  5 &  4 &  5 &  6 &  4 &  6 &  5 & 6  \\
11000 & 29.8 & 4 &   2 &  4 &  6 & 3 &   5 &  6 &  4 &  5 &  6 & 3  &  4 &  4 & 4 \\
14000 & 15.2 & 6 &   3 &  6 &  9 & 3 &   5 &  6 &  3 &  4 &  5 & 2  &  3 &  3 & 3 \\
\hline
\end{tabular}
\end{minipage}
\end{table*}

Once defined the proper setup, we proceed to present the realistic profiles to be fed to the profiler in the CL tests. Table \ref{tab:cn2prof1} shows a compilation of typical $C_n^2$ profiles obtained in Paranal from a mix of SLODAR, MASS-DIMM and SCIDAR measurements collected along several years \citep{Sarazin:2013aa}. Groups have been divided according to seeing bins in terms of `median', `bad' and `good' profiles, based on their turbulence contributions closer to the ground, i.e. those easier to be corrected by the AOF. These groups have, respectively, median values of $\overline{h}$ (equation \ref{eq:h}) of 3.73, 4.46 and 2.70 km. The 7$^\mathrm{th}$ column lists the probability of observing such profiles in this database. In the simulations both the turbulent heights and the Na layer altitudes/thicknesses (with default values of 90/7 km, respectively) are multiplied by $\sec{z}$, given that we assume here a default zenith angle of 30$\degree$. Also as input for the simulations, the reference wind profile $v_\mathrm{ref}$ has to be scaled by a constant $\beta$, given in Tables \ref{tab:cn2prof1} and \ref{tab:cn2prof2}, such that
\begin{equation}
\overline{v} = \left(\frac{\sum_h [ \beta v_\mathrm{ref}(h) ]^{5/3} C_n^2(h)}{\sum_h C_n^2(h)}\right)^{3/5}
\end{equation}
\noindent where the relation
\begin{equation}
\tau_0 = 0.314~\frac{r_0}{\overline{v}}
\end{equation} 
\noindent holds. For the seeing $S$ and coherence times $\tau_0$ shown in Table \ref{tab:cn2prof1} the $\beta$ values have been calculated and used in the Octopus simulations.

Finally, we input an outer scale of 25 m in all simulations and restrict ourselves to analyzing only normalized profiles for the reasons exposed previously. 

\subsection{GLAO Modes}
\vspace{0.5cm}
\subsubsection{GRAAL}

GRAAL is one of the GLAO or `seeing-enhancer' modes for the AOF. It is aimed to feed the 7.5'$\times$7.5' FoV imager HAWK-I on the Nasmyth A focus of UT4. Its goals are an improvement of about 40 per cent in the {\it K}-band FWHM and to achieve observations with 0.3" FWHM in $\sim$50 per cent of the time under a seeing slightly below 1" \citep{2012SPIE.8447E..38P}. As shown in Table \ref{tab:aofsystems}, the separation angle for the HR and LR baselines are 11.6' and 8.2', respectively. This gives the profiler an unprecedented resolution in probed altitudes but a maximum altitude constrained to less than 2.89 km at zenith, and 2.49 km at $z$ = 30$\degree$. Therefore, for the typical Paranal listed $C_n^2$ profiles with equivalent altitudes $\overline{h}$ ranging between 2.39 and 4.88 km, the amount of unseen turbulence can be significant, lying between 8-33 per cent for $z$ = 30$\degree$.
\begin{figure*}
\hspace{-18mm}
\includegraphics[scale=0.56]{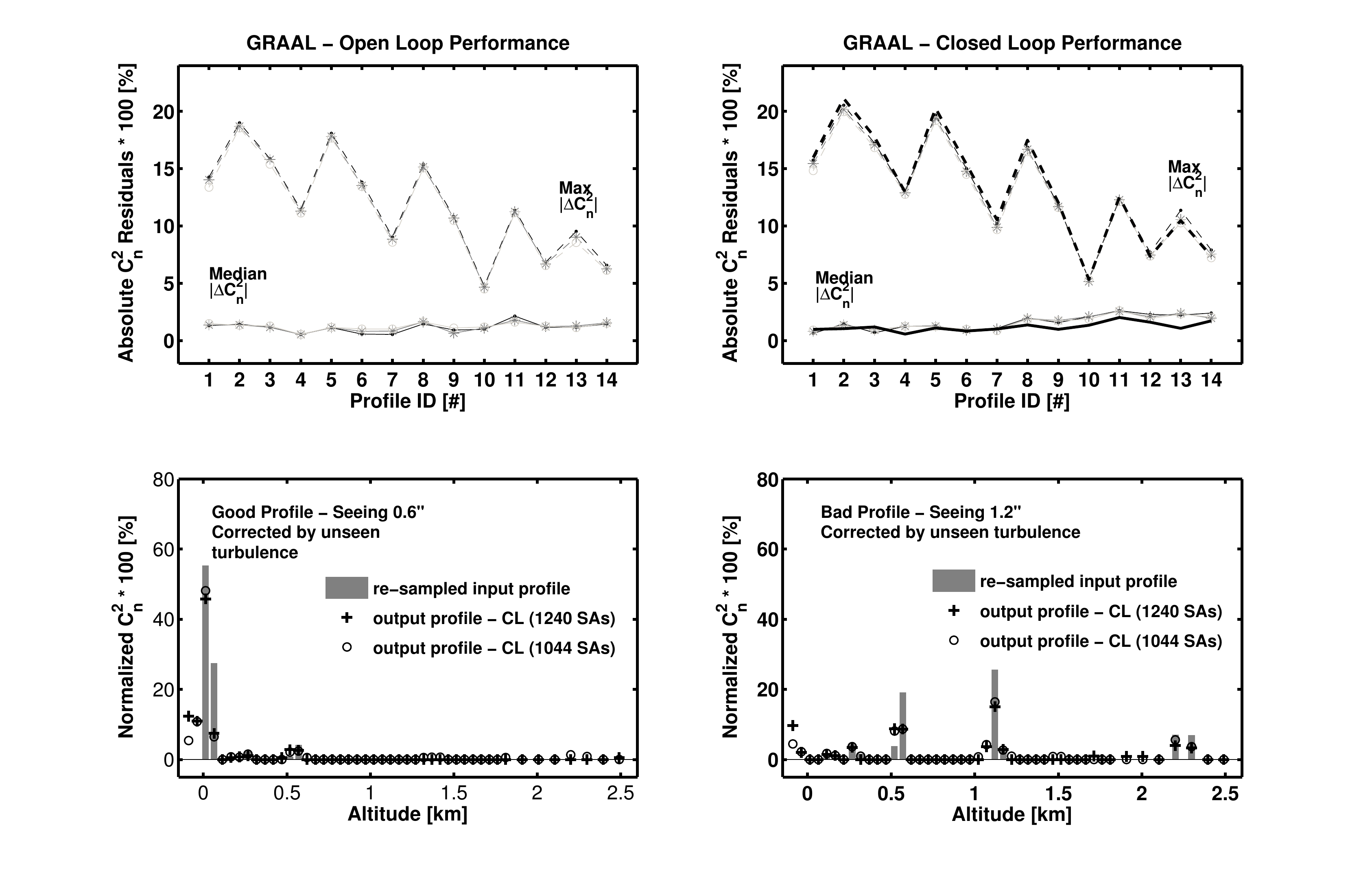}
 \caption{Upper plots: performance of the profiler for GRAAL with the 14 typical Paranal profiles in OL (left) and CL (right). Solid lines: median absolute deviations; dashed lines: maximum absolute deviations; dots: $\alpha=0$ noise filtering; stars: $\alpha=1$ noise filtering; open circles: $\alpha=2$ noise filtering. Thick lines in the top-right plot refer to results obtained using the mask of only 1044 SAs and $\alpha=1$ noise filtering. Lower plots: two examples of outputs from the profiler using CL simulations: of a good profile with 0.6" seeing and 8 per cent of unseen turbulence (left) and of a bad profile with 1.2" seeing and 26 per cent of unseen turbulence (right). All output profiles have been corrected by the unseen turbulence using as a reference the input value of the simulations, $r_0^\mathrm{sim}$. Noise filtering of $\alpha=1$ applied to the output profiles in the lower plots.}
\label{fig:graalprof}
\end{figure*}

Fig. \ref{fig:graalprof} (top) shows the performance of the profiler for all the normalized profiles listed in Tables \ref{tab:cn2prof1} and \ref{tab:cn2prof2}, under different noise filters ($\alpha=0$, 1 and 2). For the sake of comparison, the results for profiles derived from OL simulations are also shown on the left. The statistics are carried out only on resampled input bins that contain turbulence. Median and maximum absolute deviations lie between $\sim$0.5-2.5 and $\sim$5-20 per cent, respectively, and do not depend significantly on the noise filter applied. There is a general trend for the error in the $C_n^2$ peak estimation to decrease for larger turbulences, in similar ways for profiles derived from either OL or CL simulations. Subsets of profiles with the same seeing tend to show a better performance (smaller maximum residuals) for a smaller turbulence content close to the ground, at least for seeing conditions better than 1.2". This is related to the trend to underestimate turbulence peak intensities, as discussed earlier. Running the profiler with the mask shown in Fig. \ref{fig:bigmask} produces no significant effect on the statistics of the residuals, as shown by the thick lines in the CL case. 

Fig. \ref{fig:graalprof} (bottom) shows two opposite examples of profiles retrieved from CL GRAAL data, a good turbulence profile with 0.6" seeing and 8 per cent of unseen turbulence, and a bad turbulence profile with 1.2" seeing and 26 per cent of unseen turbulence. The normalized output profiles were noise filtered with $\alpha=1$ and unseen turbulence corrected using $r_0^\mathrm{sim}$ as a reference; they have been retrieved using both the default and the 1044 SAs masks (denoted by crosses and open circles, respectively). The very two first bins of the GRAAL profile refer to altitudes conjugated below the ground (-89 and -38 m) but we see that, using the default mask, the contribution obtained in those bins are not negligible, amounting to $\sim$12-23 per cent in the shown cases. It is important to highlight that this effect is not seen in OL data, and therefore it is related to the DSM correction and/or to the POL reconstruction. By applying the mask of 1044 SAs we managed to reduce such amount to $\sim$6-16 per cent, with the contribution in the second altitude bin staying practically unaltered.  This might indicate that either a more restrictive mask is needed and/or that there is a cross-talk effect between turbulent bins, as seen previously. In general, despite the detected biases in the negative altitude bins we see that the profiler succeeds qualitatively in identifying the major turbulence peaks and the overall shape of the input profile. The underestimation of the output CL profile (using the default mask) in specific peaks can be improved if we set as zero the turbulence in the two first bins and re-do the normalization. For instance, in the 0.6" case, the peaks at altitudes 13 m and 62 m reduce their residuals from $\sim$10 and 20 to -3 and 18 per cent, respectively. Given all the uncertainties, errors in the retrieved unseen turbulences for all the profiles can reach 16 per cent, with an average value of 8 per cent. 

\begin{figure*}
\hspace{-18mm}
\includegraphics[scale=0.56]{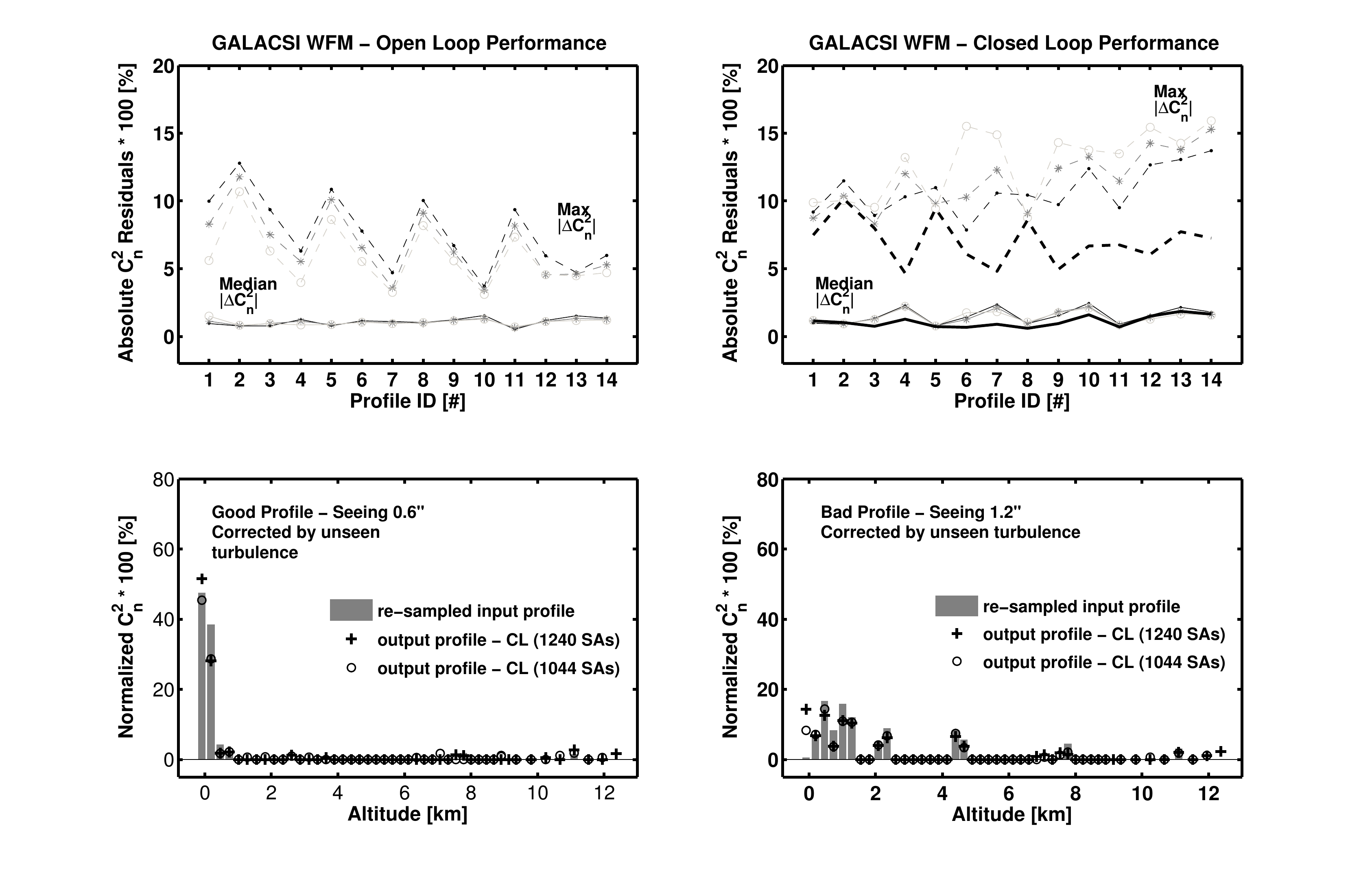}
 \caption{Upper plots: performance of the profiler for GALACSI WFM with the 14 typical Paranal profiles in OL (left) and CL (right). Solid thin lines: median absolute deviations; dashed thin lines: maximum absolute deviations; dots: $\alpha=0$ noise filtering; stars: $\alpha=1$ noise filtering; open circles: $\alpha=2$ noise filtering. Thick lines in the top-right plot refer to results obtained using the mask of only 1044 SAs and $\alpha=1$ noise filtering. Lower plots: two examples of outputs from the profiler using CL simulations: of a good profile with 0.6" seeing  (left) and of a bad profile with 1.2" seeing (right), both with only $\sim$3 per cent of input unseen turbulence. All output profiles have been corrected by the unseen turbulence using as a reference the input value of the simulations, $r_0^\mathrm{sim}$. Noise filtering of $\alpha=1$ applied to the output profiles in the lower plots.}
\label{fig:galwfmprof}
\end{figure*}

\subsubsection{GALACSI WFM}
The other AOF GLAO mode is GALACSI WFM. It is expected to double the ensquared energy within a 0.2"$\times$0.2" sky area at 750 nm with seeing conditions between 0.6" and 1.1" \citep{2012SPIE.8447E..37S}. These partially corrected wavefronts will be sent to the visible integral field spectrograph MUSE, located at the Nasmyth B focus of UT4. With the 4 LGSs pointed 64" off-axis, the MUSE science FoV intended to be corrected by this mode is of 1' across. In this mode, the amount of turbulence not detected by the profiler can be significant depending on the seeing conditions. It lies between 2-9 per cent for $z$ = 30$\degree$ for the 14 compiled profiles. At zenith, the altitudes can be probed up to 14 km and the spatial resolutions in the individual profiles lie between 252-640 m.

Fig. \ref{fig:galwfmprof} shows the performance of the profiler for this mode. Here the residuals in CL differ significantly from the reference OL ones, with an increase in maximum absolute deviations for worse turbulence conditions. Again, we see an overexcitation of the first altitude bin in CL, more or less in the same range as detected for GRAAL ($\sim$10-15 per cent), when using the 1240 SAs default mask. This effect in the first altitude bin, reflected in the maximum absolute deviation of the residuals, is significantly minimized -- down to an average of 7 per cent -- when the 1044 SAs mask is applied, in particular for profiles with larger seeings (see the dashed thick line in Fig. \ref{fig:galwfmprof}, top-right). However, and as for the GRAAL case, applying such mask does not improve the result in the second altitude bin for the best seeing profiles, this being the reason why the maximum absolute deviation of the residuals does not decrease in that regime. This might also reflect the need for a more restrictive mask, which would decrease the first peak of turbulence and increase the contribution of the second one through the renormalization of the profile. In general, the profiler applied to this mode tends to identify well all the turbulence peaks above the ground with respect to the input profile in the simulations. For the 14 profiles, the unseen turbulence is retrieved with an average and maximum errors of 5 and 12 per cent, respectively.

\vspace{0.5cm}
\subsubsection{2-Layer Profile: the $\Gamma$ Parameter}

In summary, the profiler succeeds in providing a overall good output for the AOF GLAO modes. Both provide unprecedented altitude resolution with respect to former systems (for instance, the best altitude resolution of GeMS is of about 1 km) with extremely good identification of main turbulent layers, and rather fair reproduction of their intensities. However, it is important to point out that for many applications robustness is preferred in detriment of spatial resolution. Some PSF reconstruction techniques, such as that described by \cite{2012SPIE_PSFReconstruction}, for example, only require a coarse idea of the atmospheric turbulence distribution (2 or 3 layers). Thus, as an exercise, we can define a quantity $\Gamma_h$ as the ratio between the cumulated $C_n^2$ up to a certain altitude $h$ and the total integrated turbulence. Since the integration of $C_n^2$ minimizes the overestimation of the intensities in the first two bins, for this test the profiler was run with the default mask.

Running a scan through possible values of $h$ we have verified that there is no optimal altitude to compute $\Gamma_h$ in order to minimize the residuals from the simulations. Therefore, {\it ad doc} $h$ values of 300  and 500 m are adopted for GRAAL and GALACSI WFM, respectively.  The resulting range of $\Gamma_{300}$ (GRAAL)  is $\sim$0.06-0.86 whereas of $\Gamma_{500}$ (GALACSI WFM) is 0.15-0.88.  The correspondence between input (from simulations) and output (from the profiler) $\Gamma_{h}$ parameters is good, with a typical rms error of $\sim$0.07 for both modes. 

\begin{figure}
\hspace{0.25cm}
\includegraphics[scale=0.76]{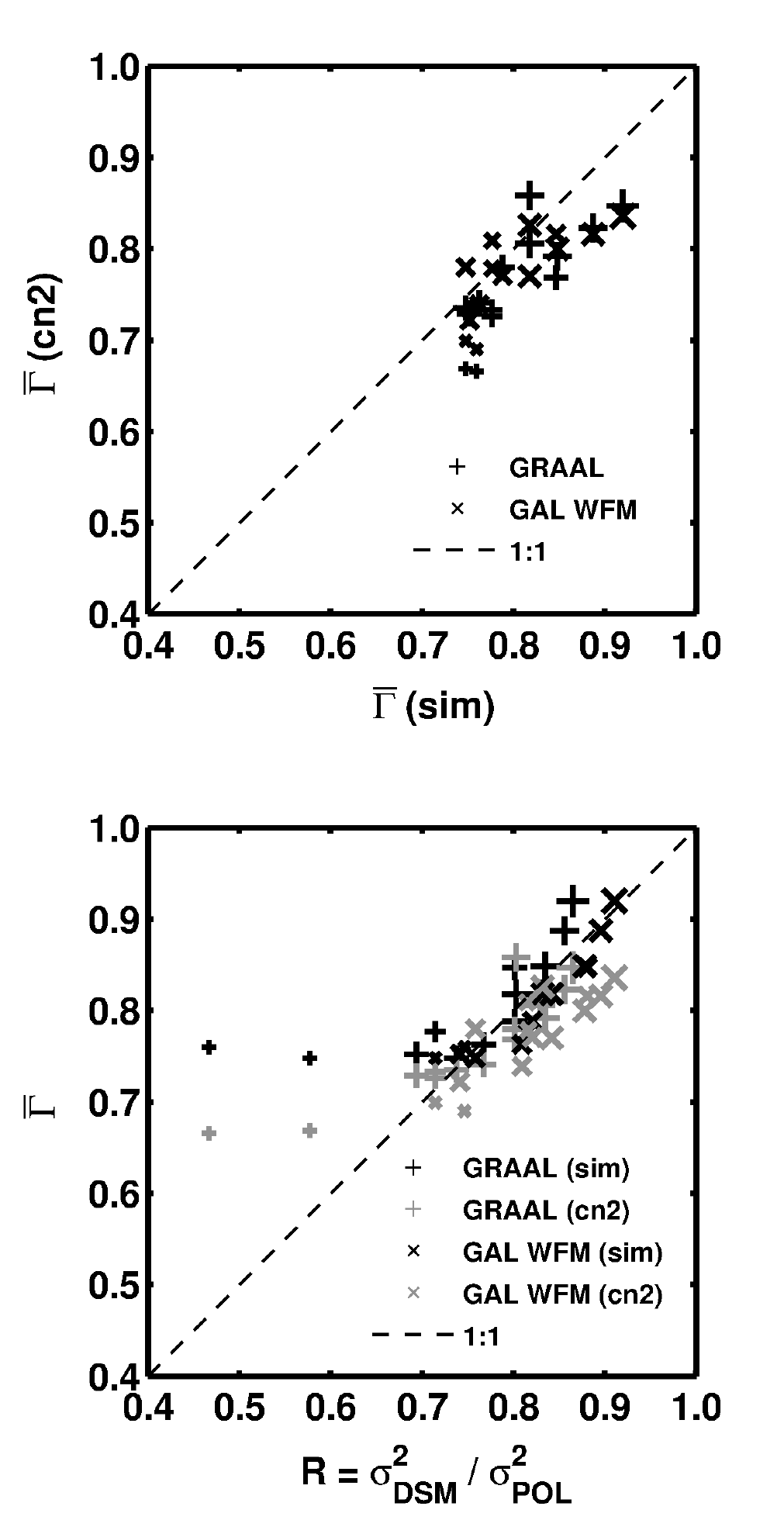}
 \caption{Top: the relation of input $\overline{\Gamma}$ from simulations versus output $\overline{\Gamma}$ from the profiler, for the AOF GLAO modes. Bottom: $\overline{\Gamma}$ versus the amount of turbulence corrected by the DSM, $R$. The scales of the symbols reflect the amount of turbulence contained in the first simulated layer (at 30 m), discretized in 20 per cent bins. Both GLAO modes discard the null-correlation hypothesis with a probability of less than 1 per cent. }
\label{fig:gammavsR}
\end{figure}

It was interesting to know if the parameter $\Gamma_h$, as defined in the previous paragraph, could be related to the total amount of the turbulence corrected by the DSM. For that purpose we computed the Zernike coefficients from the POL slopes and also from the DSM commands (via slopes space), using a proper slopes-to-Zernikes projection matrix with 150 modes. Possible measurement errors were subtracted using the auto-correlation method by \cite{1464-4258-6-6-014}. Neglecting piston and TT components, we integrated the error-corrected variances of all remaining Zernike modes for both POL and DSM-only components. The ratio of the DSM variance to the POL variance (called hereafter $R$) was then computed for each profile simulation, resulting in a range of 0.47-0.87 for GRAAL and of 0.71-0.91 for GALACSI WFM. It turns out that there is indeed a correlation of $\Gamma_h$ with $R$ for both modes, but rather steep, not providing any information about the altitude until which the turbulence is corrected in the GLAO modes. 

An alternative approach consisted in comparing $R$ with $\overline{\Gamma} = \Gamma_{h'}$, with $h' = \mbox{min}(\overline{h}, h_\mathrm{max})$, where $\overline{h}$ is the already defined equivalent altitude. Fig. \ref{fig:gammavsR} (top) shows how the input (`sim') and output (`cn2') $\overline{\Gamma}$ correlate for the two AOF GLAO modes, in CL regime. Notice that for GRAAL only two profiles have $\overline{h}$ lying inside the probed range of the profiler; so for all the others we integrate the whole profile obtained -- properly corrected by the unseen turbulence -- to obtain $\overline{\Gamma}$\footnote{Given the discretization of the simulated profiles used here this ends up being equivalent to integrating the turbulence up to $\overline{h}$.}. Despite the larger scattering in the results and the smaller range probed by this new parameter (0.75-0.92), input and output $\overline{\Gamma}$ are fairly well correlated (null-hypothesis of correlation with a probability of $\la$0.2 per cent). The parameter $\overline{\Gamma}$ obtained from the profiler tends to be slightly underestimated, with a robust fit giving an angular coefficient of 0.86 and 0.64 for GRAAL and GALACSI WFM, respectively. Assuming a 1:1 relation, the residual rms for both AOF modes lies between 0.04-0.06. 

The parameter $\overline{\Gamma}$ correlates very well with $R$ (bottom of Fig. \ref{fig:gammavsR}). An exception occurs for the GRAAL mode with profiles that have a very low content of turbulence at the ground (profiles \#10 and \#13, both with less than 20 per cent in the first simulated altitude layer at 30 m). Discarding these two profiles for GRAAL, all subsets shown in this specific plot have a null-hypothesis of correlation between $\overline{\Gamma}$ and $R$ below 0.1 per cent.  Moreover, assuming a 1:1 relation, the rms of the residuals lies between 0.03-0.06. In summary, $R$ seems to be a good indicator of the correction performed by the DSM up to the average altitude $\overline{h}$, as long as $R \ga 0.6$. Also, the parameter $\overline{\Gamma}$ provides a better criterion for comparing input and output profiles, in particular a criterion closer to quantities that matter on the real system (amount of turbulence corrected by the AO).

\subsection{LTAO Mode}
\begin{figure*}
\hspace{-18mm}
\includegraphics[scale=0.56]{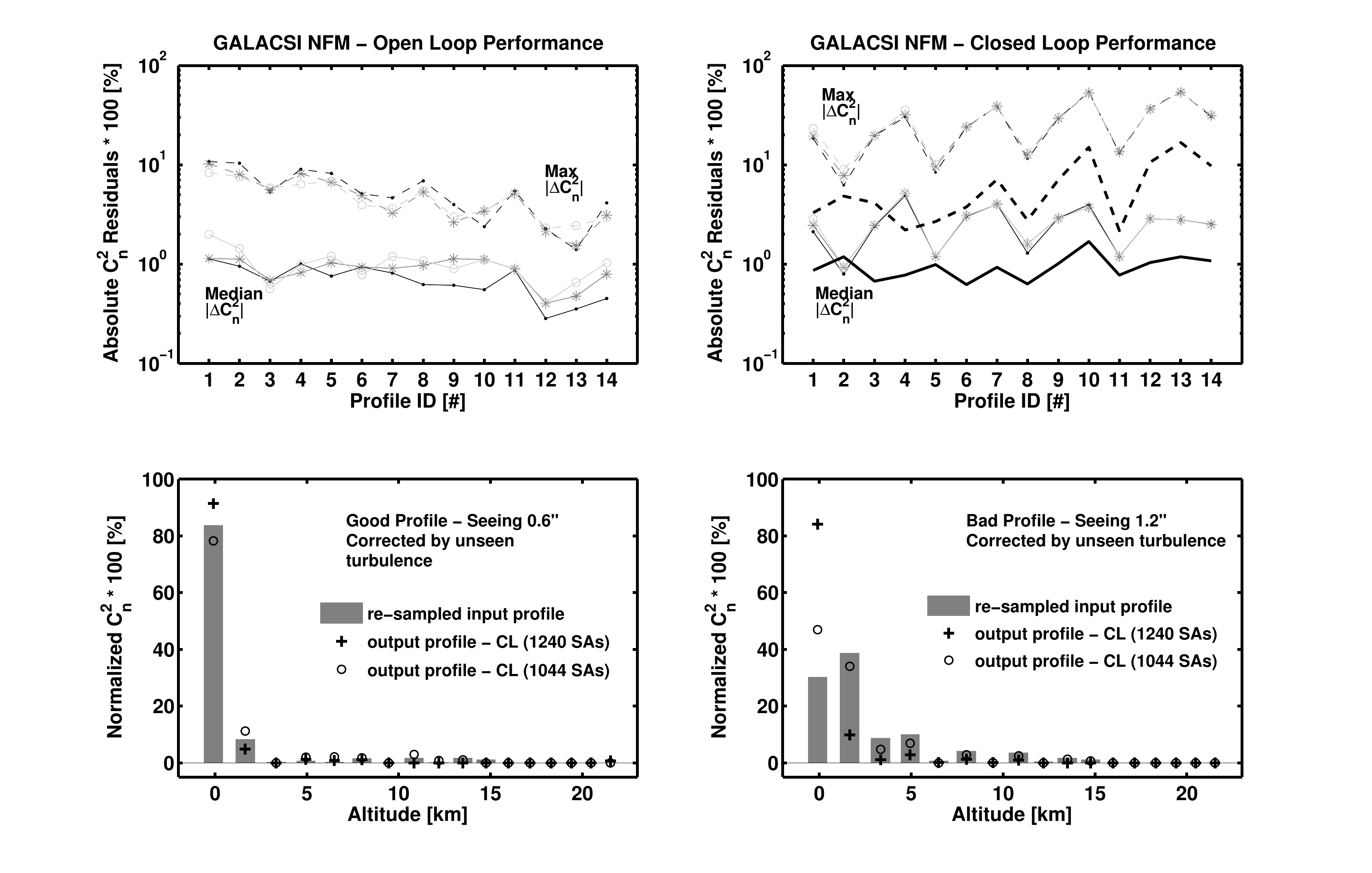}
 \caption{Upper plots: performance of the profiler for GALACSI NFM with the 14 typical Paranal profiles in OL (left) and CL (right). Solid lines: median absolute deviations; dashed lines: maximum absolute deviations; dots: $\alpha=$0 noise filtering; stars: $\alpha=$1 noise filtering; open circles: $\alpha=$2 noise filtering. Thick lines in the top-right plot refer to results obtained using the mask of only 1044 SAs and $\alpha=$1 noise filtering. Lower plots: two examples of outputs from the profiler using CL simulations: of a good profile with 0.6" seeing  (left) and of a bad profile with 1.2" seeing (right). Noise filtering of $\alpha=1$ applied to the output profiles in the lower plots.}
\label{fig:galnfmprof}
\end{figure*}

GALACSI NFM is the LTAO mode for AOF. This mode aims to optimize the correction for the center of the FoV -- of 20" diameter -- and to provide a high correction. This translates into a Strehl larger than 5 per cent -- with a goal of 10 per cent -- at 650 nm for a 0.6" seeing and using a NIR TT source of at least 15 mag in {\it J-H} bands - \cite{2012SPIE.8447E..37S}. 

The main utility of the profiler for this mode is the ability to locate the most important turbulent layers, useful for optimizing its performance. The reconstruction process for GALACSI NFM uses a virtual DM approach, described in \cite{2004MNRAS.349.1009L}. The interaction matrix is calculated using several virtual deformable mirrors conjugated to different heights (2 virtual DMs were used in these simulations, conjugated to 0 and 3000 m). These DMs have the same number and geometry of actuators as the deformable secondary. Using a propagation operator, the cone effect is taken into account. Once a synthetic interaction matrix is created with all the DMs (virtual and real), it is inverted using a MAP algorithm. After that, all actuator commands from all the DMs are projected onto the DSM. This operation is done by simply adding -- actuator by actuator -- all the commands.

Fig. \ref{fig:galnfmprof} shows the spatial resolution provided by the HR profile, repeating the analysis conducted for other modes. Up to 14 km (the maximum height of the simulated profiles) about 10 layers can be sampled, and at least 30 per cent of the turbulence -- considering the 14 typical profiles -- will be concentrated in the first output bin. The figure shows that there is a significant loss of the profiler performance  when using the default mask; the minimum turbulence detected in the first bin, for all profiles, stays always above 81 per cent. A noticeable improvement in the residuals is verified when using the 1044 SAs mask: the median and maximum absolute deviations of the residuals drop from an average of $\sim$2.8 and 20 to $\sim$1 and 6 per cent, respectively. Even with the final configuration for the LTAO mask not yet defined, awaiting further improvements on the loop correction, this test has shown that by properly masking out problematic SAs one can improve significantly the estimation of the $C_n^2$ profile.   

\subsection{Final Remarks}

It has been verified that for the 3 AOF modes -- GRAAL, GALACSI WFM \& GALACSI NFM -- it is possible to characterize relatively well the turbulence distribution with the chosen $C_n^2$ profiler once a correlation mask is properly defined. The importance of this mask seems to increase with the level of correction expected from the mode, being crucial for LTAO. The major peaks of the input $C_n^2$ distribution are identified in all modes, and quantitatively, at least for the 1044 SAs mask chosen here, the median absolute deviations of the profile residuals lie within $\sim$2.5 per cent. There is still some intensity cross-talk between the very first bins, specially in the best resolved case -- GRAAL -- which has to be further addressed if such a fine resolution is actually needed; this cross-talk is the major cause of the large maximum absolute deviations of the residuals observed for this mode (up to 20 per cent). 

\section[]{SUMMARY AND CONCLUSIONS}
\label{sec:discussion}

We have validated the so-called wind-profiler algorithm under the AOF configuration. This validation was carried out using end-to-end Octopus simulations. 

In a first instance, OL simulations allowed us to check the profiler limitations and to tackle its sensitivity issues in aspects which are intrinsic to the profiler, such as the effect of uncertainties in the input parameters (the outer scale and Na layer altitude), sampling effects and response to different types of turbulence distribution.  

The detected noise level in bins with simulated zero turbulence is consistent with 0.5 per cent error.
We have, however, found some leakage from the profile output intensities whenever the simulated turbulent layer is strong and falls exactly on an output bin altitude. This amount of leakage, which can reach up to 20 per cent for a single input turbulent layer, can be regulated to some extent by adopting a noise filter. When spatial resolution is not an issue, the integration of neighbouring output bins can decrease the residuals with respect to the input turbulence. Alternatively, it is also possible to retain only the LR output profile instead of the combined one. 

The absolute calibration of the output profile depends strongly on an assumption for the turbulence outer scale, a parameter which cannot be directly retrieved with the presented algorithm. A way to circumvent this problem is to get an independent estimate of turbulence parameters and to use them in the assumed atmospheric model for the profile. Alternatively, it is possible to opt for the method presented in \cite{Valenzuela2014}, which makes an estimate of the total $L_0$ by fitting the slopes autocovariances and highlights the need for a more detailed study of the outer scale altitude distribution in order to properly calibrate the $C_n^2$ profile. Bearing these calibration issues in mind and considering that the most important information for the AOF performance improvement is contained in the relative $C_n^2$ distribution, our work was focused on testing only normalized profiles. Nevertheless, Fried parameter estimates given by the profiler and by the WFSs slopes variances could still be useful for the correction of the unseen turbulence, regardless of the assumed outer scale. 

Dataset length and undersampling can have an impact on the normalized $C_n^2$ estimation. This study recommends for the AOF a minimum acquisition of 15k data frames to feed the SPARTA profiler routine, which is then submitted to an undersampling of $\tau = 10$. Preliminary tests running in a parallelized {\sc Matlab} environment have shown that the profile computation of such amount of data is viable in about 15 seconds, and therefore suitable for real time estimations.

In CL, a reliable $C_n^2$ estimation largely depends on the knowledge about the quality of correction achieved all over the pupil. This happens in particular for GALACSI NFM, and, to a lesser extent, to GALACSI WFM. The design of realistic correlation masks -- that in the future should take into account features such as obscuring spiders and possible defective actuators -- are a crucial step in order to avoid the introduction of spurious correlation signals in the profiler. The masking procedure has already been used in previous works with this profiler (e.g. for GeMS), but might need a special attention here because of the different DM nature.

Using a symmetrical {\it ad hoc} mask of 1044 SAs to analyse simulations with the 14 typical Paranal turbulence profiles, the profiler has provided good estimates for all the AOF modes. The GLAO modes are able to provide an unprecedented spatial resolution in the $C_n^2$ determination. It is recommended that unseen turbulence correction of the derived $C_n^2$ profile is carried out only for the GRAAL mode. For GALACSI modes this correction is either unnecessary (NFM) or negligible (WFM), in particular for the latter if we consider the uncertainties involved versus the amount of unseen turbulence considered. If a fine resolution for the profile is not required in the GLAO modes, it is possible to define a parameter $\overline \Gamma$, integrating the $C_n^2$ output profile up to the altitude $\overline{h}$, determined from the same profile. The parameter $\overline \Gamma$ has then a direct relation with the amount of turbulence corrected by the DSM. The profiler used with the LTAO mode succeeds in providing the locations of the main turbulence peaks with a spatial resolution within $\sim$1.0-1.5 km, enabling one to fine tune the loop reconstruction process in real time. 

\section{Acknowledgments}

This work was supported by CNPq, Conselho Nacional de Desenvolvimento Cient\'ifico e Tecnol\'ogico - Brazil. The authors thank the anonymous referee for comments and suggestions.
\bibliography{astro-ph_file}

\label{lastpage}

\end{document}